\documentclass[journal]{IEEEtran}

\ifCLASSINFOpdf
\else
\fi

\usepackage{amssymb}
\usepackage{amsmath}
\usepackage{amsfonts}
\usepackage{amssymb}
\usepackage{mathtools}
\usepackage{enumerate}
\usepackage{graphicx}
\usepackage[english]{babel}
\usepackage{multirow}
\usepackage{booktabs}
\usepackage{array}
\usepackage{todonotes}
\newcolumntype{L}[1]{>{\raggedright\let\newline\\\arraybackslash\hspace{0pt}}m{#1}}
\newcolumntype{C}[1]{>{\centering\let\newline\\\arraybackslash\hspace{0pt}}m{#1}}
\newcolumntype{R}[1]{>{\raggedleft\let\newline\\\arraybackslash\hspace{0pt}}m{#1}}

\usepackage[utf8]{inputenc}

\usepackage{color,soul}
\usepackage{color, colortbl}
\definecolor{Yellow}{rgb}{1.0,1,0}

\begin{document}

\title{Reducing the Hausdorff Distance in Medical Image Segmentation with Convolutional Neural Networks}

\author{Davood~ Karimi,
		and~Septimiu~ E.~ Salcudean,~\IEEEmembership{Fellow,~IEEE}% <-this % stops a space
\thanks{D. Karimi is with the Department
of Electrical and Computer Engineering, University of British Columbia, Vancouver,
BC, V6T 1Z4, Canada, e-mail: karimi@ece.ubc.ca}
\thanks{S. E. Salcudean is with the Department of Electrical and Computer Engineering, University of British Columbia, Vancouver, BC, Canada.}}

\maketitle

\begin{abstract}
The Hausdorff Distance (HD) is widely used in evaluating medical image segmentation methods. However, existing segmentation methods do not attempt to reduce HD directly. In this paper, we present novel loss functions for training convolutional neural network (CNN)-based segmentation methods with the goal of reducing HD directly. We propose three methods to estimate HD from the segmentation probability map produced by a CNN. One method makes use of the distance transform of the segmentation boundary. Another method is based on applying morphological erosion on the difference between the true and estimated segmentation maps. The third method works by applying circular/spherical convolution kernels of different radii on the segmentation probability maps. Based on these three methods for estimating HD, we suggest three loss functions that can be used for training to reduce HD. We use these loss functions to train CNNs for segmentation of the prostate, liver, and pancreas in ultrasound, magnetic resonance, and computed tomography images and compare the results with commonly-used loss functions. Our results show that the proposed loss functions can lead to approximately $18-45 \%$ reduction in HD without degrading other segmentation performance criteria such as the Dice similarity coefficient. The proposed loss functions can be used for training medical image segmentation methods in order to reduce the large segmentation errors.
\end{abstract}


\IEEEpeerreviewmaketitle

\section{Introduction}
\label{Introduction}

\IEEEPARstart{I}{mage} segmentation is the process of delineating an object or region of interest in an image. It is a central task in medical image analysis, where the volume of interest has to be isolated for visualization or further analysis. Some of the applications of medical image segmentation include measuring the size or shape of the volume of interest, creating image atlases, targeted treatment, and image-guided intervention.

Medical image segmentation has been the subject of numerous papers in recent decades. In many applications, the manual segmentation produced by an expert radiologist is still regarded as the gold standard. However, compared with manual segmentation, computerized semi-automatic and fully-automatic segmentation methods have the potential for increasing the speed and reproducibility of the results \mbox{\cite{toennies2017,zhou2015}}. Fully-automatic segmentation methods eliminate the inter-observer and intra-observer variability that are caused by such factors as the differences in expertise and attention and errors due to visual fatigue. Moreover, especially with the emergence of convolutional neural network (CNN)-based segmentation algorithms in recent years, great progress has been made in reducing the performance gap between automatic and manual segmentation methods \mbox{\cite{roth2018c,wang2018}}.

\begin{figure}[t]
    \centering
    \includegraphics[width=60mm]{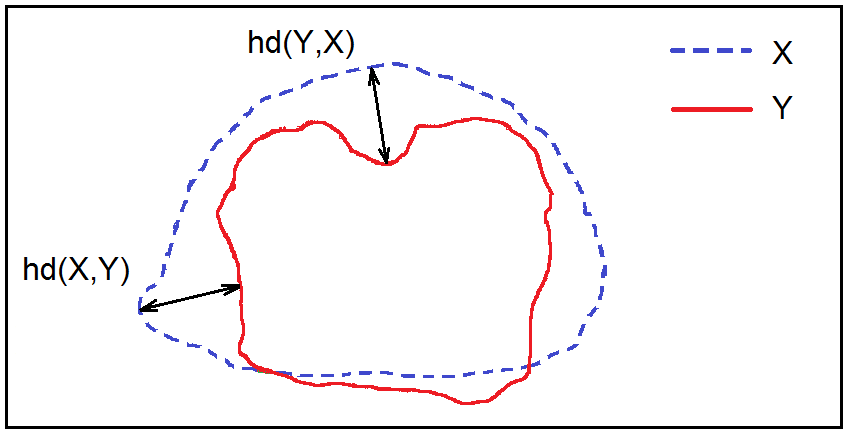}
    \caption{A schematic showing the Hausdorff Distance between points sets $X$ and $Y$.}
    \label{fig:hausdorff_schematic}
\end{figure}

The performance of automatic segmentation methods is usually evaluated by computing some common objective criteria such as the Dice similarity coefficient (DSC), mean boundary distance, volume difference or overlap, and Hausdorff Distance (HD) \cite{crum2006,taha2015b}. Among these, HD is one of the most informative and useful criteria because it is an indicator of the largest segmentation error. In some applications, segmentation is one step in a more complicated multi-step process. For example, some multimodal medical image registration methods rely on segmentation of an organ of interest in one or several images. In such applications, the largest segmentation error as quantified by HD can be a good measure of the usefulness of the segmentations for the intended task. As illustrated in Figure \ref{fig:hausdorff_schematic}, for two point sets $X$ and $Y$, the one-sided HD from $X$ to $Y$ is defined as \cite{rockafellar2009}:

\begin{equation} \label{eq:hd1}
\text{hd}(X,Y)= \adjustlimits\max_{x\in X} \min_{y\in Y} \| x -y \|_2,
\end{equation}

\noindent and similarly for $\text{hd}(Y,X)$:

\begin{equation} \label{eq:hd11}
\text{hd}(Y,X)= \adjustlimits\max_{y\in Y} \min_{x\in X} \| x -y \|_2.
\end{equation}

The bidirectional HD between these two sets is then:

\begin{equation} \label{eq:hd2}
\text{HD}(X,Y)= \text{max} \big( \text{hd}(X,Y) , \text{hd}(Y,X)  \big)
\end{equation}

In the above definitions we have used the Euclidean distance, but other metrics can be used instead. Intuitively, $\text{HD}(X,Y)$ is the longest distance one has to travel from a point in one of the two sets to its closest point in the other set. In image segmentation, HD is computed between boundaries of the estimated and ground-truth segmentations, which consist of curves in 2D and surfaces in 3D.

Although HD is used extensively in evaluating the segmentation performance, segmentation algorithms rarely aim at minimizing or reducing HD directly \mbox{\cite{smistad2015,heimann2009}}. For example in the segmentation methods based on deformable models, the typical formulation of the external energy used to drive the segmentation algorithm is an integral (i.e., sum) of the edge information along the segmentation boundary \cite{caselles1997, kass1988}. Therefore, these methods can be interpreted as minimizing the mean error over the segmentation boundary. Atlas-based segmentation methods, which are another class of widely-used techniques, work by registering a set of reference images to the target image by minimizing such global loss functions as the sum of squared difference of image intensities or the mutual information \cite{marroquin2002, park2003}. Similarly, machine learning-based image segmentation methods aim at reducing a global loss function rather than the largest segmentation error \cite{makni2014,pereira2016}.

To the best of our knowledge, with one exception \cite{schmidt2012}, no previous study has proposed a method for directly minimizing or reducing HD in medical image segmentation. There may be several reasons why previous works have not targeted HD. One reason is that unlike many other criteria such as cross-entropy, DSC, and volume overlap that are affected by the segmentation performance over the entire image, HD is determined solely by the largest error. If a segmentation algorithm is designed to focus on the largest error, the overall segmentation performance may suffer. Moreover, an algorithm that aims solely at minimizing the largest error may be unstable. This is confirmed by our own observations in this study, which are discussed in Section \mbox{\ref{Results}} of this paper. Moreover, especially for segmentation of complex structures that are common in medical imaging, a segmentation algorithm may achieve satisfying accuracy over most of the image but have large errors at one or a few isolated locations. This could occur because of different reasons such as weak or missing edges, artifacts, or low signal to noise ratio. In these cases, accurate segmentation is difficult or impossible even for a human expert. Hence, it may not be reasonable to expect high segmentation accuracy everywhere in the image because ``the ground truth" may be unreliable or nonexistent. The sensitivity of HD to noise and outliers has been well documented in the computer vision literature. For image matching, for example, it has been suggested to use modified definitions of HD or to combine HD with other image information to obtain more robust algorithms \cite{dubuisson1994,yang2007}. Two widely-used variations of HD that have been designed to reduce the sensitivity to outliers are Partial HD \cite{huttenlocher1993} and Modified HD \cite{dubuisson1994}. Partial HD replaces the $\max$ operation in Equation \eqref{eq:hd1} with the $K^{th}$ largest value, whereas Modified HD replaces it with the averaging operation.

Moreover, direct minimization of HD is very challenging from an optimization viewpoint. Most of the studies in computer vision that have used HD have focused on a restricted set of problems such as object matching or face detection. In these applications, the goal is to match a template $B$ to an image $A$ subject to simple transformations such as translation, rotation, and scaling. Hence, the goal is to find a small set of parameters $p$ such that the HD between the transformed $B$ and $A$, i.e., $\text{HD}(T_p(B),A)$, is minimized. Even this restricted scenario is not easy to handle. Some studies have suggested methods such as genetic algorithms \cite{kirchberg2002}, while others have used exhaustive search to solve the problem \cite{sim1999, tan2006}. A number of studies have proposed similar formulations for medical image registration by approximately minimizing the HD between reference and target landmarks under rigid transformations \cite{knauer2011, dimitrov2007}. We are aware of only one work that has used HD in medical image segmentation \cite{schmidt2012}. That study was quite different from the methods proposed in this work. In particular, the authors of \mbox{\cite{schmidt2012}} focus on the specific problem of multi-surface segmentation where each small surface is nested within a larger surface. Moreover, their segmentation method is based on minimizing an energy function that consists of a data term and a smoothness term. Instead of minimizing HD, they propose to use the prior knowledge about the maximum value of HD as a constraint. They show that the resulting problem could be NP-hard and suggest simplifying assumptions in order to obtain an approximate solution. They apply their method for segmentation of different structures in MR and ultrasound images. However, they only visually display their results and do not provide any quantitative evaluation. They also do not compare their method against any other methods.

The goal of this paper is to propose methods for reducing HD in CNN-based segmentation methods. CNN-based methods are relative new-comers to the field of medical image segmentation, but they have already proved to be highly versatile and effective \mbox{\cite{litjens2017,roth2018c}}. These methods usually produce a dense (i.e., pixel- or voxel-wise) segmentation probability map of the organ or volume of interest, although there are some exceptions \mbox{\cite{milletari2017,karimi2018segmentation}}. The early CNN-based image segmentation methods applied a soft-max function to the output layer activations and defined a loss function in terms of the negative log-likelihood, which is equivalent to a cross-entropy loss for binary segmentation \cite{long2015}. Later, some studies proposed different loss functions to address specific challenges of medical image segmentation. For example, some works suggested a weighted cross-entropy, where larger weights are assigned to more important regions such as the boundaries of the volume of interest \cite{ronneberger2015,anas2017}. Another difficulty in medical image segmentation is that often the object of interest occupies a small portion of the image, biasing the algorithm towards achieving higher specificity than sensitivity. To counter this effect, it was suggested that DSC be used as the objective function \cite{milletari2016}. Another study has suggested that more control over sensitivity and specificity can be achieved by using the Tversky Index as the objective function \cite{salehi2017}. These studies have shown that the choice of the loss function can have a large impact on the performance of image segmentation methods. Recently, some studies have argued that the choice of a good loss function for training of deep learning models has been unfairly neglected and that research on this topic can lead to large improvements in the performance of these models \cite{janocha2017}.

In this paper, we propose techniques for reducing HD in CNN-based medical image segmentation. The novel aspects of this work are as follows: 1) We propose three different loss functions based on HD that, to the best of our knowledge, are novel and have not been used for medical image segmentation before, 2) We use four datasets to segment different organs in different medical imaging modalities and empirically show that using these loss functions can significantly reduce large segmentation errors, 3) Through extensive experiments we show the potential benefits and challenges of using HD-based loss functions for medical image segmentation.

The paper is organized as follows. In Section \ref{MaterialsAndMethods}, we propose three methods for estimating HD from the output probability map of a CNN. Because minimizing HD directly may not be desirable and could lead to unstable training, based on each of the three methods of estimating HD we propose an ``HD-inspired" loss function that can be used for stable training. After explaining our methods in Section \ref{MaterialsAndMethods}, we will present and discuss our results on four different medical image datasets in Section \ref{Results}. We will describe the conclusions of this work in Section \ref{Conclusions}.

\section{Materials and Methods}
\label{MaterialsAndMethods}

\subsection{Notations}

\begin{figure}[!ht]
    \centering
    \includegraphics[width=50mm]{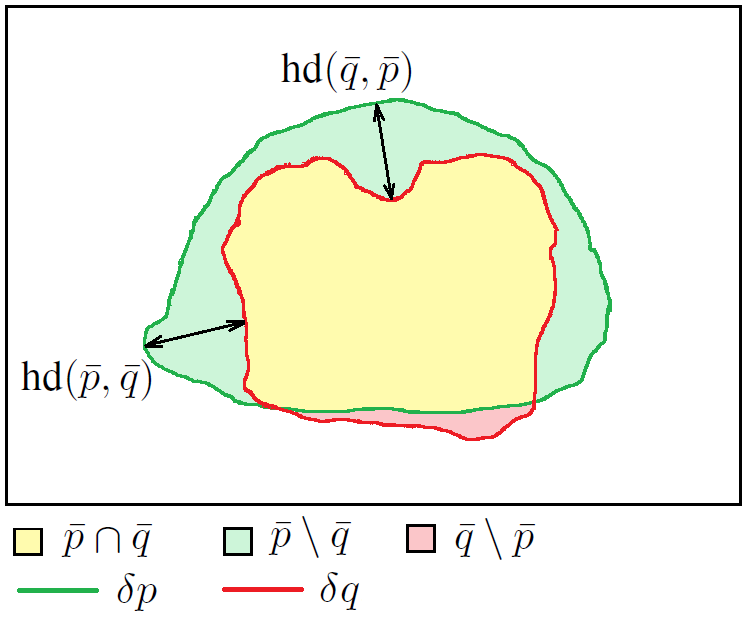}
    \caption{A visual depiction of some of the notations used in this paper.}
    \label{fig:notations_figure}
\end{figure}

We denote the output segmentation probability map of a CNN with $q \in [0,1]$. To obtain a binary segmentation of the object versus the background, one usually thresholds $q$ at 0.50 to get a segmentation map with values in $\{ 0,1 \}$, where 0 indicates the background and 1 indicates the object. We denote this binary map with $\bar{q}$. Similarly, we denote the ground-truth segmentation with $p \in [0,1]$ and $\bar{p} \in \{ 0,1 \}$, although typically the ground-truth segmentation is a binary map, i.e., $p \equiv \bar{p}$. As shown in Figure \ref{fig:notations_figure}, we denote the boundaries of $\bar{p}$ and $\bar{q}$ with $\delta p $ and $\delta q$, respectively. For ease of illustration, in this section we use 2D figures to explain our proposed methods. However, the extension of the methods to 3D is trivial, and we will present experimental results with 2D as well as 3D images in Section \ref{Results}.  

\subsection{Estimation of the Hausdorff Distance Based on Distance Transforms}
\label{DistanceTransform}

Our first approximation of HD is based on distance transforms (DT). The DT of a digital image is a derived representation of that image where each pixel has a value equal to its distance to an object of interest in the image. For a 2D binary image $X[i,j]$, with 0 representing the background and 1 indicating the object, we have \cite{szeliski2010}:

\begin{equation} \label{eq:dt0}
\text{DT}_{\text{X}}[i,j]= \min_{k,l; X[k,l]=1} d \big( [i,j], [k,l] \big)
\end{equation} 

\noindent where $d$ denotes the distance between pixel locations and $[k,l]$ denote the indices of the object (i.e., foreground) pixels.. In this work, we use the standard choice of the Euclidean distance: $d \big( [i,j], [k,l] \big) = \sqrt{(k-i)^2+(l-j)^2}$.

\begin{figure}
  \centering
\begin{tabular}{c c c}
    \includegraphics[width=25mm]{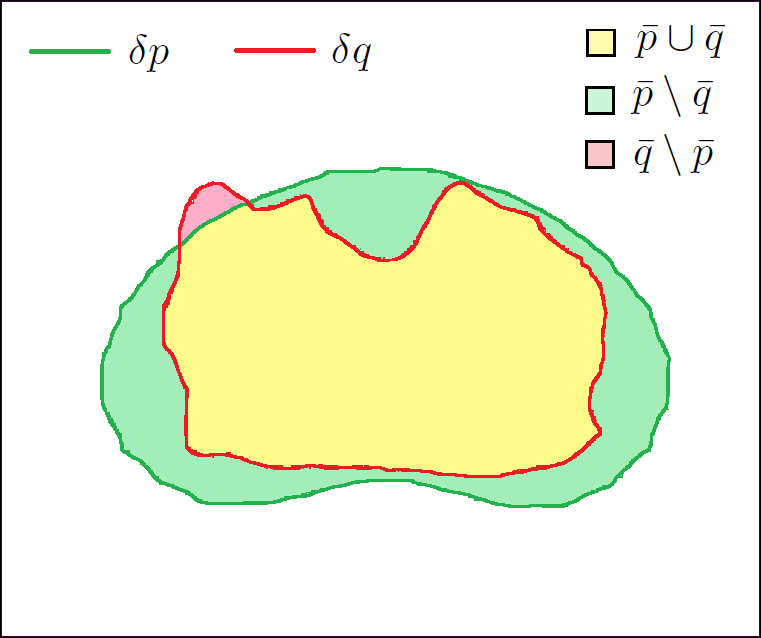} & 
    \includegraphics[width=25mm]{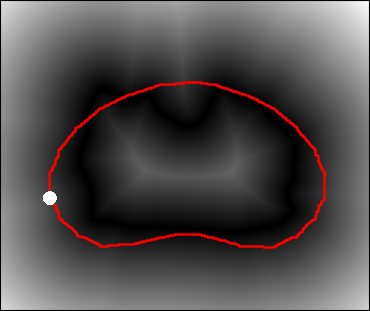}  &
    \includegraphics[width=25mm]{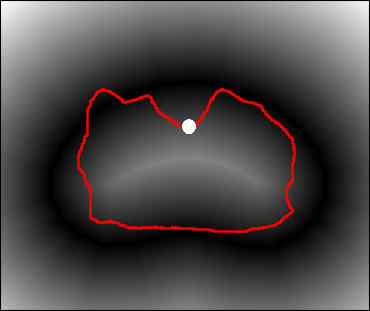} \\ 
    (a) & (b) & (c)  \\
  \end{tabular}
\caption{(a) An example of a 2D ground-truth and predicted segmentations, denoted with $\bar{p}$ and $\bar{q}$, respectively. (b) The distance transform $d_q$ with $\delta p$ overlaid in red. The white circle shows the location of the largest $d_q$, which corresponds to $\text{hd}(\delta p,\delta q)$. (c) Similar to (b), for finding $\text{hd}(\delta q,\delta p)$.}
\label{fig:distance_transform_schematic}
\end{figure}

Here, we define the distance map of the ground-truth segmentation as the unsigned distance to the boundary, $\delta p$, and denote it with $d_p$. Similarly, $d_q$ is defined as the distance to $\delta q$. As shown in Figure \ref{fig:distance_transform_schematic}, it is clear that we can write:

\begin{equation} \label{eq:dt1}
\text{hd}_{\text{DT}}(\delta q,\delta p)= \max_{\Omega} \big( ( \bar{p} \bigtriangleup \bar{q} ) \circ  d_p  \big)
\end{equation} 

\noindent where we have used the subscript ${\text{DT}}$ to indicate that HD is computed using the distance transforms. In the above equation, and in the rest of this paper, $\bigtriangleup$ denotes the set operation of symmetric difference defined as $\bar{p} \bigtriangleup \bar{q}= (\bar{p} \setminus \bar{q}) \cup (\bar{q} \setminus \bar{p})$ \cite{jech2002}. For us, this can be simply computed as $\bar{p} \bigtriangleup \bar{q}= \mid \bar{p} - \bar{q} \mid$. Moreover, in the above equation, $\circ$ denotes the Hadamard (i.e., entry-wise) product and $\Omega$ denotes the grid on which the image is defined, which means that $\text{max}$ is with respect to all pixels.

We can similarly compute $\text{hd}_{\text{DT}}(\delta p,\delta q)$, and then $\text{HD}_{\text{DT}}(\delta q,\delta p)$ as follows:

\begin{equation} \label{eq:dt2}
\text{hd}_{\text{DT}}(\delta p,\delta q)= \max_{\Omega} \big( ( \bar{p} \bigtriangleup \bar{q} ) \circ  d_q  \big)
\end{equation}

\begin{equation} \label{eq:dt3}
\text{HD}_{\text{DT}}(\delta q,\delta p)= \max \big( \text{hd}_{\text{DT}}(\delta q,\delta p), \text{hd}_{\text{DT}}(\delta p,\delta q) \big)
\end{equation}

\begin{figure}[!ht]
\centering
\begin{tabular}{c c}
    \includegraphics[width=40mm]{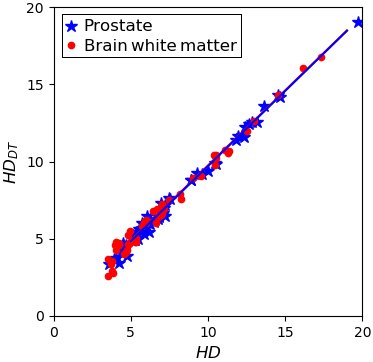} & 
    \includegraphics[width=43mm]{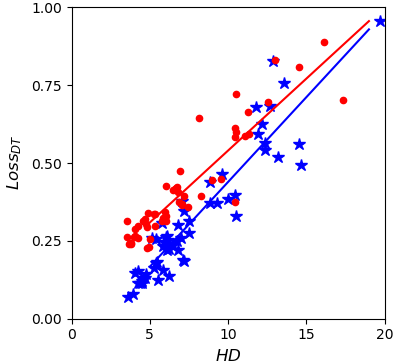} \\ 
    (a) & (b)  \\
\end{tabular}
\caption{Plots of $\text{HD}_{\text{DT}}$ and $\text{Loss}_{\text{DT}}$ versus exact HD for sets of 3D MR prostate and brain white matter images and their rough segmentations.}
\label{fig:distance_transform_proof}
\end{figure}

Figure \ref{fig:distance_transform_proof}(a) illustrates that this method is a correct way of computing HD. In this figure, we have plotted the HD estimated using Equation \mbox{\eqref{eq:dt3}} versus the exact HD computed using \mbox{\cite{taha2015}} between the ground-truth and approximate segmentations of 50 3D Magnetic Resonance (MR) prostate images and 50 brain white matter MR images. For both prostate and brain data the Pearson correlation coefficient of the fitted linear function is above 0.99. Based on the above estimation of HD, we suggest the following loss function for CNN training:

\begin{equation} \label{eq:dt_cost}
\text{Loss}_{\text{DT}}( q, p)= \frac{1}{\mid \Omega \mid} \sum_{\Omega} \Big(  (p-q)^2  \, \circ  ( d_p^{\alpha} + d_q^{\alpha} ) \Big)
\end{equation}

Compared with Equation \eqref{eq:dt3} that is an accurate estimator of HD, this loss function is different in three aspects. First, instead of focusing only on the largest segmentation error, we smoothly penalize larger segmentation errors. The parameter $\alpha$ determines how strongly we penalize larger errors. To determine a good value for this parameter, we tried values of $\alpha \in \{ 0.5, 1.0, ..., 3.5, 4.0 \}$ in small cross-validation experiments. Our experiments showed that values of $\alpha$ between 1.0 and 3.0 led to good results. In all the experiments reported for this method in this paper, we used a value of $\alpha=2.0$, which we found to be the best value in that set. Second, unlike Equation \eqref{eq:dt3}, we use $p$ and $q$ instead of the thresholded maps, $\bar{p}$ and $\bar{q}$, to allow the training to take into account this useful information. Finally, instead of $\mid p-q \mid$, we use $(p-q)^2$. This choice is inspired by the results of \cite{janocha2017} and will be justified empirically in Section \ref{Results}.

Figure \mbox{\ref{fig:distance_transform_proof}(b)} shows a plot of $\text{Loss}_{\text{DT}}$ versus exact HD for the same prostate and brain MR data as in \mbox{\ref{fig:distance_transform_proof}(a)}. The loss functions have been scaled to $[0,1]$ for display. For both prostate and brain data the Pearson correlation coefficient for the fitted linear function is approximately 0.93. It is clear that there is a strong correlation between the two, so that reducing $\text{Loss}_{\text{DT}}$ should lead to a decrease in HD.

A drawback of this method is the high computational cost of computing the distance transforms, $d_p$ and $d_q$. In this work, we used the algorithm proposed in \cite{felzenszwalb2004} for computing DT in 2D and the algorithm in \cite{maurer2003} for experiments with 3D images. One may use less accurate but faster algorithms instead, because very accurate estimation of DT is not needed in this application. Nonetheless, the computational cost will remain high, especially in 3D. Moreover, the cost will be much higher for computing $d_q$ than for $d_p$. This is because $q$ changes during training and therefore $d_q$ should be re-computed for all images after each training epoch. On the other hand, $d_p$ needs to be computed only once. Therefore, one way of reducing the computational cost is to only consider the one-sided HD, $\text{hd}_{\text{DT}}(\delta q,\delta p)$. This leads to the following modified loss function (where we use OS to indicate ``one-sided"):

\begin{equation} \label{eq:dt_cost_os}
\text{Loss}_{\text{DT-OS}}( q, p)= \frac{1}{\mid \Omega \mid} \sum_{\Omega} \big( (p-q)^2 \circ \, d_p^{\alpha} \big)
\end{equation}

We will present some experimental results with this loss function as well in Section \ref{Results}.

\subsection{Estimation of the Hausdorff Distance Using Morphological Operations}
\label{MorphologicalOperations}

\begin{figure}[!ht]
  \centering
\begin{tabular}{c c}
    \includegraphics[width=40mm]{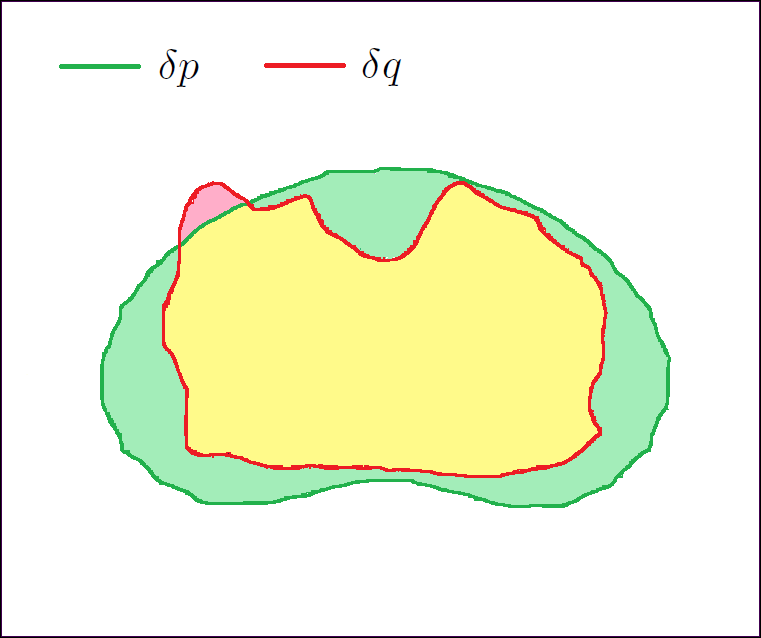} & 
    \includegraphics[width=40mm]{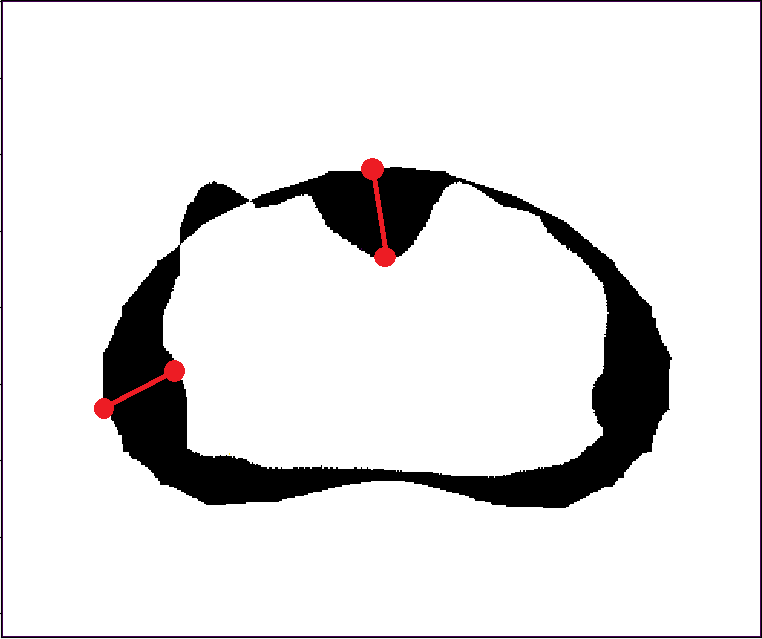} \\
    (a) & (b) \\
    \includegraphics[width=40mm]{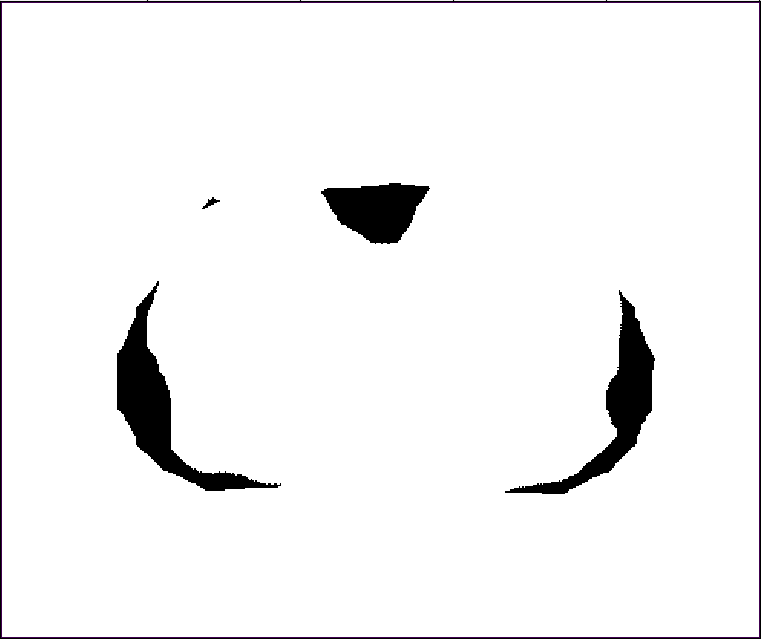} & 
    \includegraphics[width=40mm]{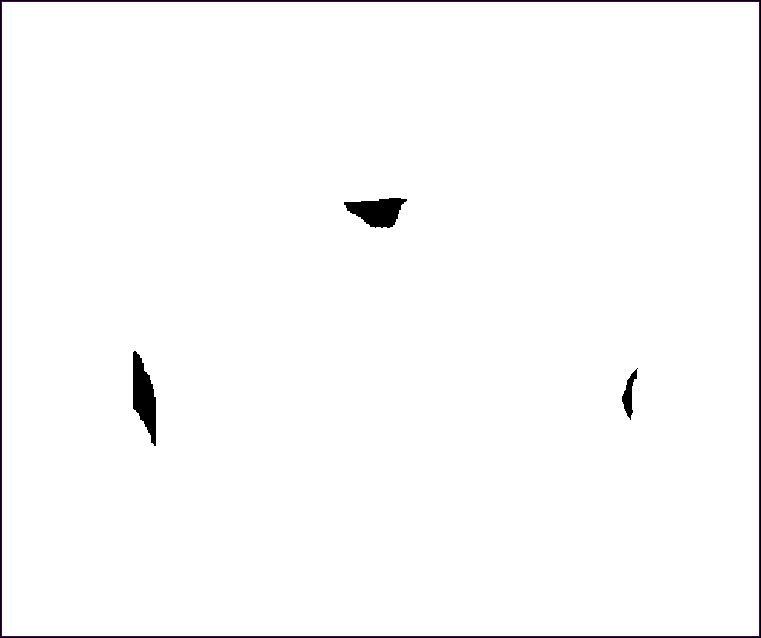} \\ 
    (c) & (d) \\
  \end{tabular}
\caption{(a) An example of a 2D ground-truth and predicted segmentations, denoted with $\bar{p}$ and $\bar{q}$, respectively. (b) The map of $\bar{p} \bigtriangleup \bar{q}$ for this example; $\text{hd}(\delta p,\delta q)$ and $\text{hd}(\delta q,\delta p)$ have been marked with red line segments on this figure. (c) and (d) The eroded map of $\bar{p} \bigtriangleup \bar{q}$ after applying, respectively, 5 and 10 erosions with a cross-shaped structuring element of size 5.}
\label{fig:morphological_operations_schematic}
\end{figure}

Although the distance transform-based method explained above is simple and intuitive, it has a high computational cost. In this section, we propose an alternative approach that is based on the use of morphological operations. As can be seen in Figure \ref{fig:morphological_operations_schematic}, $\text{HD}(\delta q,\delta p)$ is roughly related to the largest thickness of the difference between the true and estimated segmentations, $\bar{p} \bigtriangleup \bar{q}$. Therefore, one can obtain an approximate estimation of $\text{HD}(\delta q,\delta p)$ by applying morphological erosion on $\bar{p} \bigtriangleup \bar{q}$. 

Morphological erosion of a binary object $S$ defined on a grid $\Omega$ using a structuring element $B$ is defined as \cite{szeliski2010}:

\begin{equation} \label{eq:erosion}
S \ominus B= \{ z \in \Omega | B(z) \subseteq S   \}
\end{equation} 

\noindent where $B(z)$ is the structuring element shifted on the grid such that it is centered on $z$.

Let us denote a structuring element with radius $r$ as $B_r$. We suggest the following approximation to $\text{HD}(\delta q,\delta p)$ based on a morphological erosion of $\bar{p} \bigtriangleup \bar{q}$.

\begin{equation} \label{eq:er1}
\begin{aligned}
\text{HD}_{\text{ER}} & (\delta q,\delta p)=  2 r^* \\
  & \hspace{15mm} \text{where} \hspace{2mm} r^*=\text{minimum} \; r \hspace{2mm}  \\ 
  & \hspace{25mm} \text{such that} \hspace{2mm} (\bar{p} \bigtriangleup \bar{q}) \ominus B_r = \emptyset \\
\end{aligned}
\end{equation}

\noindent where the subscript $\text{ER}$ indicates that the Hausdorff Distance is computed using morphological erosion.

\begin{figure}
  \centering
\begin{tabular}{c c}
    \includegraphics[width=40mm]{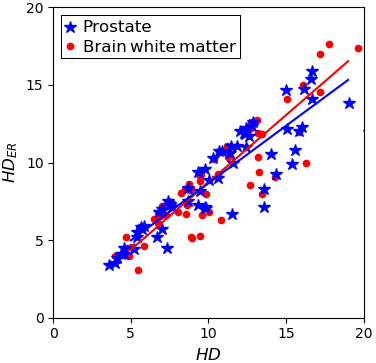} & 
    \includegraphics[width=42mm]{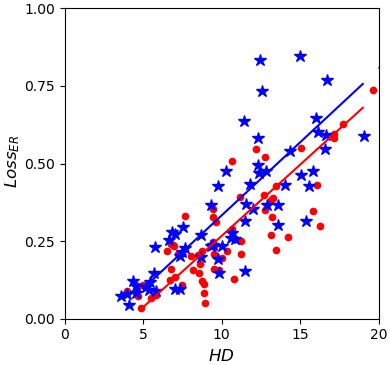} \\ 
    (a) & (b) \\
  \end{tabular}
\caption{Plots of $\text{HD}_{\text{ER}}$ and $\text{Loss}_{\text{ER}}$ versus exact HD for sets of 3D MR prostate and brain white matter images and their rough segmentations.}
\label{fig:morphological_operations_proof}
\end{figure}

$\text{HD}_{\text{ER}}$ defined above is a lower bound on the true HD because if erosion of $\bar{p} \bigtriangleup \bar{q}$ with $B_{r^*}$ does not result in an empty set then $\text{HD}>r^*$. One can make up pathological examples for which HD is much larger than $\text{HD}_{\text{ER}}$. However, as can be seen in Figure \ref{fig:morphological_operations_proof}(a), in practice the proposed  $\text{HD}_{\text{ER}}$ is a good approximation of the exact HD. The Pearson correlation coefficient for the fitted linear function in this figure for the prostate and brain data is 0.93 and 0.91, respectively.

As can be seen by comparing Figures \ref{fig:distance_transform_proof}(a) and \ref{fig:morphological_operations_proof}(a), $\text{HD}_{\text{ER}}$ is not as accurate as $\text{HD}_{\text{DT}}$. However, an advantage of $\text{HD}_{\text{ER}}$ is that morphological operations can be implemented efficiently using convolutional operations and thresholding. This had been demonstrated long before the recent surge of interest in CNNs \cite{mazille1989, razaz1999}. Therefore, $\text{HD}_{\text{ER}}$ can be computed more efficiently than $\text{HD}_{\text{DT}}$. Similar to what we did for $\text{HD}_{\text{DT}}$ above, instead of using $\text{HD}_{\text{ER}}$ directly as the loss function, we propose the following relaxed loss function:

\begin{equation} \label{eq:er_cost}
\text{Loss}_{\text{ER}}( q, p)= \frac{1}{\mid \Omega \mid} \sum_{k=1}^K \sum_{\Omega} \big( (p-q)^2 \ominus_k B \big) k^{\alpha}
\end{equation}

In the above equation we have used $\ominus_k$ to denote $k$ successive erosions. Note that erosion is applied to $(p-q)^2$, which is not binary. This is based on the generalized definition of erosion proposed in \cite{mazille1989}. In our work, we compute this via convolution with a kernel whose elements sum to one followed by a soft thresholding \cite{foucart2013} at 0.50. For 2D, we use a cross-shaped structuring element $B= \begin{pmatrix} 0 & 1/5 & 0 \\ 1/5 & 1/5 & 1/5 \\ 0 & 1/5 & 0 \end{pmatrix}$. Similarly, for 3D we use a convolutional kernel of size 3 with the center element and its 6-neighbors set to $1/7$ and the remaining 20 elements set to zero. In Equation \eqref{eq:er_cost}, $K$ denotes the total number of erosions. Increasing $K$ will increase the computational cost. On the other hand, $K$ should be large enough because all parts of $\bar{p} \bigtriangleup \bar{q}$ that remain after $K$ erosions will be weighted equally. In practice, one has to set $K$ based on the expected range of segmentation errors. We set $K=10$ for all experiments in this work. The parameter $\alpha$ determines how strongly we penalize larger segmentation errors. We used $\alpha=2.0$ in our experiments. Similar to the method in Section \mbox{\ref{DistanceTransform}}, we chose this value using a grid search.

Figure \ref{fig:morphological_operations_proof}(b) shows a plot of $\text{Loss}_{\text{ER}}$ versus the exact HD on a set of images. There is a good correlation between $\text{Loss}_{\text{ER}}$ and HD. The Pearson correlation coefficient for the fitted linear function in this figure for both prostate and brain data is 0.83. One can easily compute $\text{Loss}_{\text{ER}}$ by stacking $K$ convolutional layers to the end of any CNN.

\subsection{Estimation of Hausdorff Distance using Convolutions with Circular/Spherical Kernels}
\label{CircularConvolutions}

\begin{figure}[!ht]
    \centering
    \includegraphics[width=55mm]{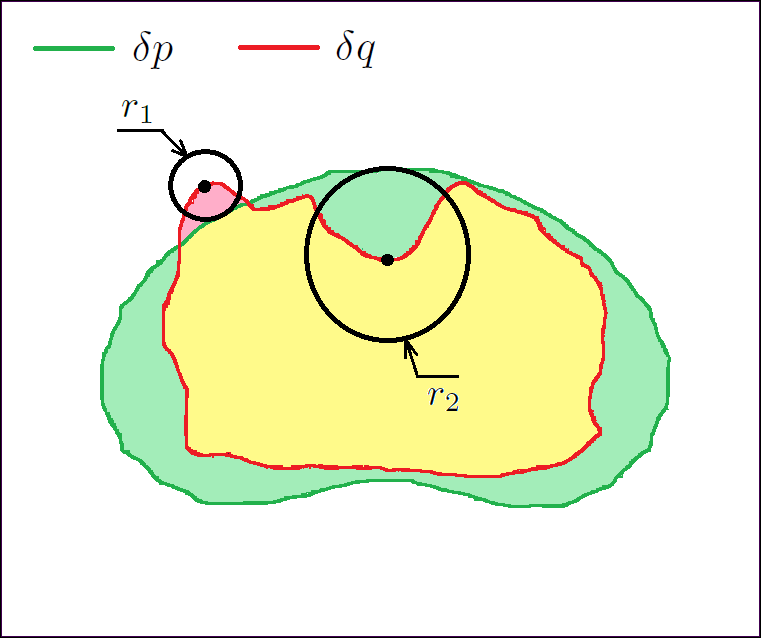}
    \caption{A schematic illustration of the method to compute HD using convolutions with circular kernels. Circles with radii $r_1$ and $r_2$ show the convolutional kernels that determine HD according to Equation \mbox{\eqref{eq:cv1}}. From this figure and Equation \eqref{eq:cv1} one can see that $\text{hd}_{\text{CV}} (\delta q,  \delta p) = \max (r_1, r_2)$.}
    \label{fig:spherical_convolution_schematic}
\end{figure}

Let us denote a circular-shaped convolutional kernel of radius $r$ with $B_r$. Elements of $B_r$ are normalized such that they sum to one. Then we can write:

\begin{equation} \label{eq:cv1}
\begin{aligned}
\text{hd}_{\text{CV}} & (\delta q, \delta p) = \max (r_1, r_2) \\
  & \text{where} \hspace{2mm} r_1=  \max r   \hspace{2mm}  \\ 
  & \hspace{5mm} \text{such that} \hspace{2mm} \max_{\Omega} f_h(\bar{p}^C \ast B_r) \circ ( \bar{q} \setminus \bar{p} ) > 0 \\
  & \text{and} \hspace{2mm} r_2=  \max r   \hspace{2mm}  \\ 
  & \hspace{5mm} \text{such that} \hspace{2mm} \max_{\Omega} f_h(\bar{p} \ast B_r) \circ ( \bar{p} \setminus \bar{q} ) > 0 \\
\end{aligned}
\end{equation}

\noindent where we have used the subscript $\text{CV}$ to denote the Hausdorff Distance computed using convolutions. In the above equation,  $\bar{p}^C= 1- \bar{p}$ denotes the complement of $\bar{p}$, and $f_h$ is a hard thresholding function that sets all values below 1 to zero. The schematic in Figure \mbox{\ref{fig:spherical_convolution_schematic}} helps the reader understand this equation. It shows that HD can be computed using only convolution and thresholding operations. We can compute $\text{hd}_{\text{CV}}(\delta p,\delta q)$ using a similar equation and then $\text{HD}_{\text{CV}}(\delta p,\delta q) = \max \big( \text{hd}_{\text{CV}}(\delta q,\delta p), \text{hd}_{\text{CV}}(\delta p,\delta q) \big)$.

\begin{figure}[!ht]
  \centering
\begin{tabular}{c c}
    \includegraphics[width=40mm]{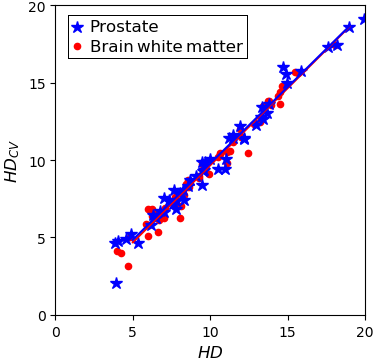} & 
    \includegraphics[width=42mm]{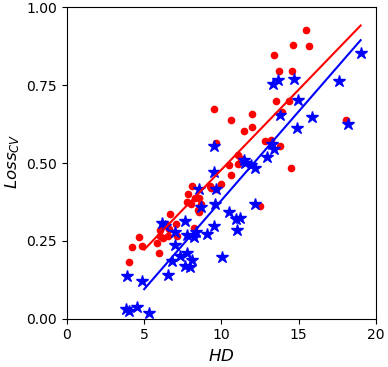} \\ 
    (a) & (b) \\
  \end{tabular}
\caption{Plots of $\text{HD}_{\text{CV}}$ and $\text{Loss}_{\text{CV}}$ versus exact HD for sets of 3D MR prostate and brain white matter images and their rough segmentations.}
\label{fig:spherical_convolution_proof}
\end{figure}

As shown in Figure \ref{fig:spherical_convolution_proof}(a), $\text{HD}_{\text{CV}}$ is an accurate approximation of the true HD. The Pearson correlation coefficient for the fitted linear function in this figure for both prostate and brain data is approximately 0.99.

We should note that Equation \mbox{\eqref{eq:cv1}} provides an exact estimate of HD in the continuous domain. However, on a discrete grid, the precision of HD computation is limited by the pixel size. Moreover, there is some discretization error when representing circular/spherical convolutional kernels on a discrete grid. This error is larger for smaller circles/spheres. As a result, the spread from the straight line is greater for smaller HD values in Figure \mbox{\ref{fig:spherical_convolution_proof}(a).}

Once again, we aim for a relaxed loss function that smoothly penalizes larger errors instead of focusing only on the largest error. Therefore, we suggest the following loss function:

\begin{equation} \label{eq:cv_cost}
\begin{aligned}
\text{Loss}_{\text{CV}}( q, p)= \frac{1}{\mid \Omega \mid} \sum_{r \in R} r^{\alpha}  \sum_{\Omega}  \big[ & f_s(B_r \ast \bar{p}^C) \circ f_{ \bar{q} \setminus \bar{p} }  + \\
& f_s(B_r \ast \bar{p}) \circ f_{ \bar{p} \setminus \bar{q} }  \ \ + \\
& f_s(B_r \ast \bar{q}^C) \circ f_{ \bar{p} \setminus \bar{q} }  + \\
& f_s(B_r \ast \bar{q}) \circ f_{ \bar{q} \setminus \bar{p} } \big] \\
\end{aligned}
\end{equation}

\noindent where $f_{ \bar{q} \setminus \bar{p} }$ is a relaxed estimation of  $\bar{q} \setminus \bar{p}$ defined as:

\begin{equation} \label{eq:relaxed_set_difference}
f_{ \bar{q} \setminus \bar{p} } = (p-q)^2  q
\end{equation}

\noindent and similarly for $f_{ \bar{p} \setminus \bar{q} }$. Moreover, we have replaced the hard thresholding, $f_h$, in Equation \eqref{eq:cv1} with the soft thresholding $f_s$ in Equation \eqref{eq:cv_cost}. The parameter $\alpha$ plays the same role here as it did in Equations \eqref{eq:dt_cost_os} and \eqref{eq:er_cost}. Similar to the above two methods, we chose $\alpha= 2.0$ via a grid search in the range $[0.50, 4.0]$. Figure \mbox{\ref{fig:spherical_convolution_proof}(b)} shows $\text{Loss}_{\text{CV}}$ as a function of the exact HD. The Pearson correlation coefficient for the fitted linear function in this figure for the prostate and brain data is approximately 0.91 and 0.88, respectively.

Similar to $\text{Loss}_{\text{ER}}$, $\text{Loss}_{\text{CV}}$ is based on convolution and thresholding operations, which can be implemented easily in deep learning software frameworks. A comparison of Figures \ref{fig:morphological_operations_proof}(a) and \ref{fig:spherical_convolution_proof}(a) shows that $\text{HD}_{\text{CV}}$ is a more accurate estimation of HD than $\text{HD}_{\text{ER}}$ is. However, whereas $\text{HD}_{\text{ER}}$ is computed using small fixed convolutional kernels (of size 3, in our implementation), computation of $\text{Loss}_{\text{CV}}$ will require applying filters of increasing size. For very large convolutional filters, especially in 3D, the computational load can become very significant. Therefore, in computing $\text{Loss}_{\text{CV}}$ we use a maximum kernel radius of 18 pixels in 2D and 9 voxels in 3D. Hence, larger segmentation errors are treated equally. To further reduce the cost, we do not use kernels of every size, but only at steps of 3, because there is no need for such fine resolution. Therefore, in Equation \eqref{eq:cv_cost} we set $R= \{ 3, 6, \dots 18 \}$ in our experiments with 2D images and $R= \{ 3, 6, 9 \}$ in experiments with 3D images. In practice, one can choose $R$ based on the expected range of segmentation errors and, of course, the pixel/voxel size.

\subsection{Data}

We used four datasets, one of 2D images and three of 3D images, in our experiments. A brief description of the data is provided below.

\subsubsection{2D ultrasound images of prostate}

This dataset consisted of trans-rectal ultrasound (TRUS) images of 675 patients. From each patient, between 7 and 14 2D TRUS images of size $415 \times 490$ pixels with a pixel size of $0.15 \times 0.15 \, \text{mm}^2$ had been acquired. The clinical target volume (CTV) had been delineated in each slice by experienced radiation oncologists using a semi-automatic segmentation software described in \cite{mahdavi2011}. The use of this software biased the segmentation of the prostate at the base and apex. The ``ground-truth" segmentation of the base and apex is also made unreliable due to the lack of clear landmarks. Therefore, we chose to work with the slices that belonged to the mid-gland, which we defined as the middle $40 \%$ of the prostate. As a result, we had a total of 1805 2D images from 450 patients for training and 820 images from the remaining 225 patients for test.

\subsubsection{3D MR images of prostate}

A total of 80 training and 30 test images were included in this dataset. This included the training data from the PROMISE12 challenge \mbox{\cite{litjens2014b}} as well as the Medical Segmentation Decathlon challenge (https://decathlon.grand-challenge.org/). The liver and pancreas datasets described below were also obtained through the Medical Segmentation Decathlon challenge. The pre-processing applied on prostate MR images included bias correction \cite{tustison2010}, resampling to an isotropic voxel size of $1 \text{ mm}^3$, and cropping to a size of $128 \times 128 \times 96$ voxels.

\subsubsection{3D CT images of liver}

This dataset consisted of 131 CT images. We used 100 images for training and 31 for test. As pre-processing, we linearly mapped the voxel values (in Hounsfield Units) from $[-1000, 1000]$ to $[0,1]$, cropping voxels smaller than -1000 and larger than 1000. We then resampled the images to an isotropic voxel size of $2 \text{ mm}^3$, and cropped them to a size of $192 \times 192 \times 128$ voxels.

\subsubsection{3D CT images of pancreas}

This dataset consisted of 282 CT images. We used 200 images for training and 82 for test. The pre-processing was similar to that for the liver CT images described above.

\subsection{CNN Architecture and Training Procedures}

Currently, the most common loss functions for CNN-based medical image segmentation are the cross-entropy and DSC. Our experience shows that, compared with cross-entropy, DSC consistently leads to better results. Therefore, we will compare the following loss functions in our experiments:

\begin{itemize}

\item DSC, defined as:

\begin{equation} \label{eq:dsc}
f_{\text{DSC}}(q, p)= 1 - \frac{2 \sum_{\Omega} (p \circ q) }{ \sum_{\Omega} (p^2 + q^2) }
\end{equation}

\item Three different HD-based loss functions defined as:

\begin{equation} \label{eq:hd_loss_general}
f^*_{\text{HD}}(q, p)= \text{Loss}_*( q, p) + \lambda \Bigg( 1 - \frac{2 \sum_{\Omega} (p \circ q) }{ \sum_{\Omega} (p^2 + q^2) } \Bigg)
\end{equation}

\noindent where $*$ is replaced with $\text{DT}$, $\text{ER}$, and $\text{CV}$ to give three different \textit{HD-based loss functions} $f^{\text{DT}}_{\text{HD}}(q, p)$, $f^{\text{ER}}_{\text{HD}}(q, p)$, and $f^{\text{CV}}_{\text{HD}}(q, p)$ based on the three losses in Equations \eqref{eq:dt_cost}, \eqref{eq:er_cost}, and \eqref{eq:cv_cost}, respectively.

As can be seen in Equation \eqref{eq:hd_loss_general}, we augment our HD-based loss term with a DSC loss term. This results in a more stable training, especially at the start of the training. We choose $\lambda$ such that equal weights are given to the HD-based and DSC loss terms. Specifically, after each training epoch, we compute the HD-based and DSC loss terms on the training data and set $\lambda$ (for the next epoch) as the ratio of the mean of the HD-based loss term to the mean of the DSC loss term. This simple empirical approach ensured that both loss terms were given equal weight and it worked well in all of our experiments.

\end{itemize}

Because the goal of this work was to study the impact of the loss function, we decided to use a standard CNN architecture and training procedure. This allows us to reduce the impact of other factors that may confound the results. Hence, we used the U-net \cite{ronneberger2015} and 3D U-net \cite{cciccek2016} for our experiments with 2D and 3D images, respectively. Data augmentation is very common in training deep learning models, especially when the training data is small. We used three standard data augmentation methods: 1) adding random noise to images, 2) random cropping, 3) random elastic deformations \cite{milletari2016,ronneberger2015}. On each of the four datasets, we used the same data augmentation parameters (e.g., the noise standard deviation and parameters of elastic deformation) for all loss functions. 

All loss functions were minimized using the Adam optimizer \mbox{\cite{kingma2014}}. We used the default parameter settings suggested in \cite{kingma2014}. The only parameter that we tuned was the learning rate because its optimal value could depend on the loss function. To conduct a fair comparison of different loss functions, for evaluating each loss function on each dataset we performed the training for 10 learning rates logarithmically spaced in $[10^{-3}, 10^{-5}]$ for 50 epochs and chose the learning rate that achieved the lowest loss on the training data. The selected learning rate was typically in the range $[10^{-3}, 10^{-4}]$. After selecting the learning rate, the model was trained from scratch for a total of 100 epochs with the selected learning rate. The learning rate was divided by 2 whenever the training loss did not decrease by more than $1 \%$ in a training epoch.

All the models and training code were implemented in Python 3.6 and TensorFlow 1.2 and run in Linux. For the exhaustive search to select the learning rate we used an NVIDIA DGX1. For training the final models (upon choosing the learning rates) we used an NVIDIA GeForce GTX TITAN X so that the reported training times be relevant to the type of hardware that most researchers currently use.

\section{Results and Discussion}
\label{Results}

Table \ref{tab:overall_summary} shows a summary of the results on the four datasets. We have used DSC and HD as the evaluation criteria. In addition to the mean and standard deviation of DSC and HD, we also report the 90th-percentile and maximum of HD on the test data. Moreover, average symmetric surface distance (ASD) values have been reported in the same table. In the last column of the same table, we have shown the training times. For DSC and HD, we performed paired t-tests on the test images to see if the results for different loss functions were significantly different. The results of these statistical tests have been shown using superscripts in this table; for each dataset, different superscripts indicate statistically significant difference at $p=0.01$. As an example, for the 2D prostate ultrasound data these superscripts indicate that: 1) In terms of DSC, all loss functions are statistically similar, 2) In terms of HD, $f^{\text{DT}}_{\text{HD}}$ and $f^{\text{CV}}_{\text{HD}}$ are statistically similar and statistically different (lower HD) than both $f^{\text{ER}}_{\text{HD}}$ and $f_{\text{DSC}}$; moreover, $f^{\text{ER}}_{\text{HD}}$ is statistically different (lower HD) than $f_{\text{DSC}}$.

\begin{table*}
\caption{Summary of the results of our experiments with four datasets. For each dataset, mean $\pm$ standard deviation of DSC, HD, and ASD are presented in addition to the maximum of DSC, 90th-percentile and maximum of HD, and the training time. Superscripts on the values of DSC and HD indicate the results of paired t-tests. For each dataset, different superscripts on DSC and HD indicate statistically significant difference at $p=0.01$.}
\label{tab:overall_summary}
\begin{tabular}{p{2.8cm} L{1.5cm} C{2.2cm} C{1.0cm} C{1.8cm} C{1.6cm} C{1.2cm}  C{1.3cm} C{1.0cm}}
\hline
 Dataset & Loss Function & DSC & Maximum of DSC & HD (mm) & 90th-percentile of HD (mm) & Maximum of HD (mm) & ASD & Training time (h) \\ \hline 
\multirow{4}{*}{2D prostate ultrasound } & 
$f_{\text{DSC}}(q, p)$               & $0.932 \pm  0.039^{\text{ a}}$ &  $0.975 $ & $4.3 \pm  2.8^{\text{ a}}$ & $7.5$ & $14.4$ &  $1.52 \pm 0.90$  & \boldmath$3.2$ \\
& $f^{\text{DT}}_{\text{HD}}(q, p)$   & \boldmath$0.946 \pm  0.041^{\text{ a}}$ & $0.982 $ &  \boldmath$2.6 \pm  1.8^{\text{ b}}$ & \boldmath$4.4$ & \boldmath$7.1$ &  \mbox{\boldmath$1.05 \pm 0.46$}  & 4.0 \\
& $f^{\text{ER}}_{\text{HD}}(q, p)$   & $0.936 \pm  0.041^{\text{ a}}$ & $\textbf{0.985} $ &  $3.0 \pm  2.0^{\text{ c}}$ & $5.6$ & $14.9$ &  $1.12 \pm 0.59$   & 3.7 \\
& $f^{\text{CV}}_{\text{HD}}(q, p)$   & $0.941 \pm  0.036^{\text{ a}}$ & $0.977 $ &  $2.7 \pm  1.8^{\text{ b}}$ & $4.5$ & $8.2$ &  $1.10 \pm 0.50$   & 4.1 \\  \hline
\multirow{4}{*}{3D prostate MRI } & 
$f_{\text{DSC}}(q, p)$                & $0.868 \pm  0.046^{\text{ a}}$ & $0.925 $ &  $7.5 \pm  3.1^{\text{ a}}$ & $10.5$ & $15.1$ & $1.95 \pm 0.46$  & \boldmath$8.0$ \\
& $f^{\text{DT}}_{\text{HD}}(q, p)$   & $0.875 \pm  0.042^{\text{ a}}$ & $\textbf{0.927} $ &  \boldmath$5.8 \pm  2.2^{\text{ b}}$ & \boldmath$7.6$ & $9.0$ & $ 1.51 \pm 0.27 $   & 21 \\
& $f^{\text{ER}}_{\text{HD}}(q, p)$   & $0.858 \pm  0.046^{\text{ a}}$ & $0.921 $ &  $6.1 \pm  2.3^{\text{ b}}$ & $8.9$ & $10.1$ &  $1.55 \pm 0.33$   & 9.8 \\
& $f^{\text{CV}}_{\text{HD}}(q, p)$   & \boldmath$0.876 \pm  0.040^{\text{ a}}$ & $0.923 $ &  \boldmath$5.8 \pm  2.5^{\text{ b}}$ & $8.0$ & \boldmath$8.4$ &  \mbox{\boldmath$1.49 \pm 0.27$}  & 14 \\  \hline
\multirow{4}{*}{3D Liver CT } & 
$f_{\text{DSC}}(q, p)$                & $0.921 \pm  0.048^{\text{ a}}$ & $0.949 $ &  $46.8 \pm  18.9^{\text{ a}}$ & $59.8$ & $72.9$ & $1.58 \pm 0.49$  & \boldmath$22$ \\
& $f^{\text{DT}}_{\text{HD}}(q, p)$   & \boldmath$0.940 \pm  0.040^{\text{ a}}$ & $0.959 $ &  \boldmath$25.1 \pm  10.3^{\text{ b}}$ & \boldmath$38.4$ & \boldmath$41.2$ & \mbox{\boldmath$1.37 \pm 0.39$}   & 47 \\
& $f^{\text{ER}}_{\text{HD}}(q, p)$   & $0.936 \pm  0.040^{\text{ a}}$ & $0.956 $ &  $31.6 \pm  12.1^{\text{ c}}$ & $44.4$ & $50.2$ & $1.44 \pm 0.41$   & 26 \\
& $f^{\text{CV}}_{\text{HD}}(q, p)$   & $0.935 \pm  0.039^{\text{ a}}$ & $\textbf{0.966} $ &  $27.3 \pm  13.4^{\text{ b}}$ & $41.8$ & $43.8$ & $1.38 \pm 0.41$   & 42 \\   \hline
\multirow{4}{*}{3D Pancreas CT } & 
$f_{\text{DSC}}(q, p)$                & $0.752 \pm  0.120^{\text{ a}}$ & $0.855 $ &  $32.1 \pm  17.0^{\text{ a}}$ & $58.5$ & $65.1$ & $2.09 \pm 0.57$   & \boldmath$22$ \\
& $f^{\text{DT}}_{\text{HD}}(q, p)$   & \boldmath$0.784 \pm  0.059^{\text{ a}}$ & $\textbf{0.870} $ &  \boldmath$21.3 \pm  11.3^{\text{ b}}$ & \boldmath$35.2$ & \boldmath$37.7$ &  \mbox{\boldmath$1.84 \pm 0.44$}  & 50 \\
& $f^{\text{ER}}_{\text{HD}}(q, p)$   & $0.767 \pm  0.066^{\text{ a}}$ & $0.845 $ &  $27.1 \pm  13.6^{\text{ c}}$ & $41.6$ & $45.0$ & $1.98 \pm 0.43$   & 24 \\
& $f^{\text{CV}}_{\text{HD}}(q, p)$   & $0.780 \pm  0.055^{\text{ a}}$ & $0.862 $ &  $21.7 \pm  11.0^{\text{ b}}$ & $39.0$ & $44.1$ & $1.91 \pm 0.39$   & 34\\
\hline
\end{tabular}
\end{table*}

On all four datasets, the three HD-based loss functions have resulted in lower average HD on the test images. The reduction in the mean of HD ranges between $18 \%$ and $45 \%$. In all cases, this reduction in HD is statistically significant. Also, with a few exceptions, the HD-based loss functions also reduced the maximum and 90th-percentile of HD on the test images. In many cases this reduction is as high as $30-50 \%$. On the other hand, the paired t-tests did not show any significant differences in terms of DSC achieved by the proposed HD-based loss functions and the pure DSC loss. This summary clearly demonstrates that the proposed loss functions effectively reduce HD in CNN-based image segmentation.

\begin{figure*}[!ht]
  \centering
\begin{tabular}{c c c c c}
	\multirow{2}{*}{ 2D prostate ultrasound}  &
    \includegraphics[width=25mm]{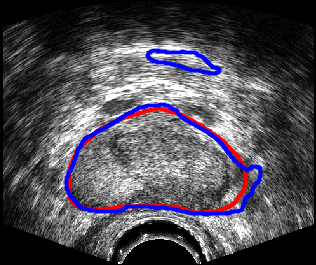} & 
    \includegraphics[width=25mm]{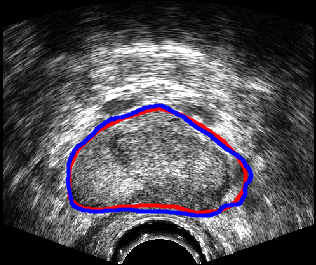} &
    \includegraphics[width=25mm]{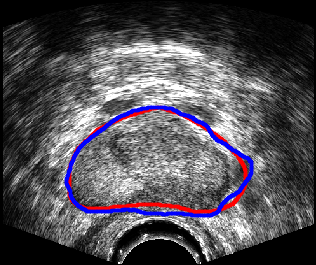} &
    \includegraphics[width=25mm]{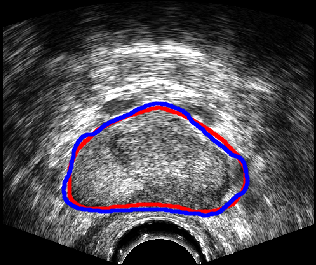} \\
    & \includegraphics[width=25mm]{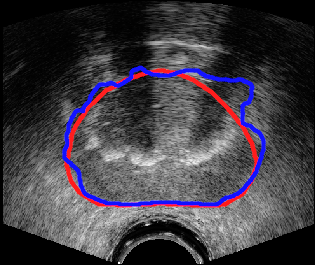} & 
    \includegraphics[width=25mm]{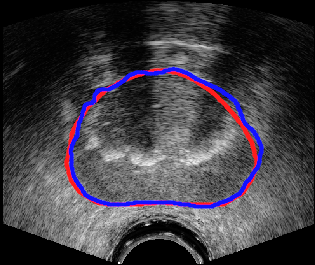} &
    \includegraphics[width=25mm]{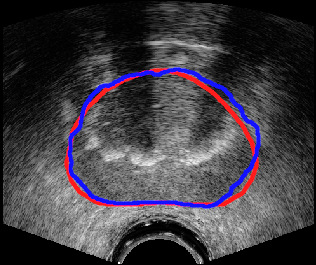} &
    \includegraphics[width=25mm]{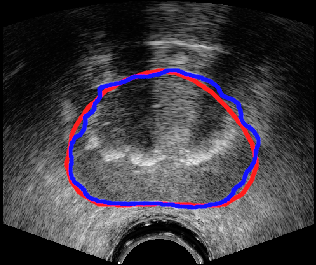} \\
    \multirow{2}{*}{ 3D prostate MRI} &
    \includegraphics[width=25mm]{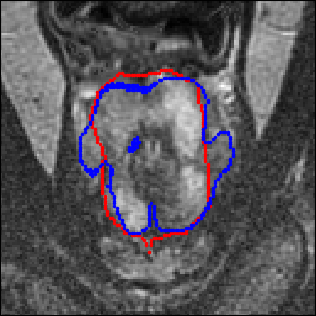} & 
    \includegraphics[width=25mm]{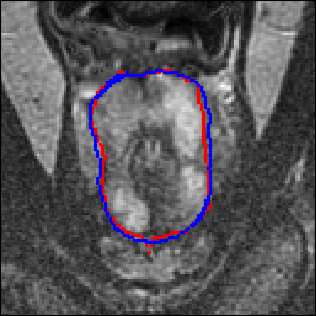} &
    \includegraphics[width=25mm]{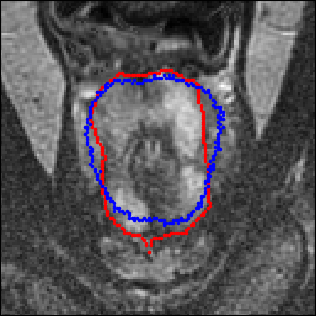} &
    \includegraphics[width=25mm]{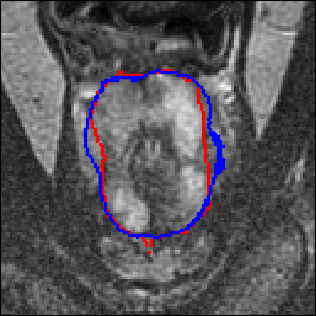} \\
    & \includegraphics[width=25mm]{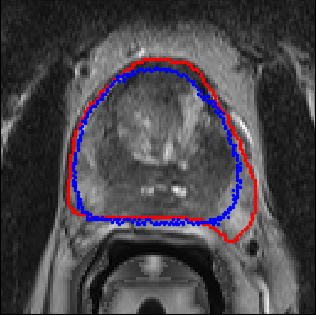} & 
    \includegraphics[width=25mm]{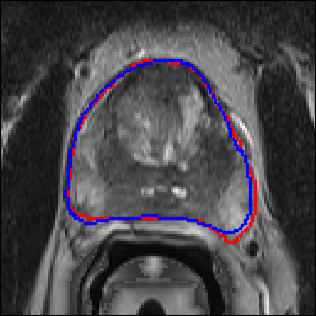} &
    \includegraphics[width=25mm]{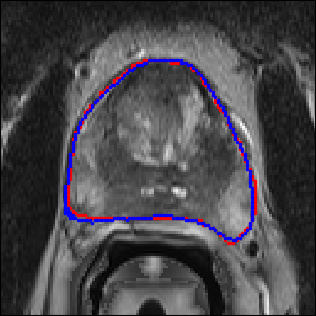} &
    \includegraphics[width=25mm]{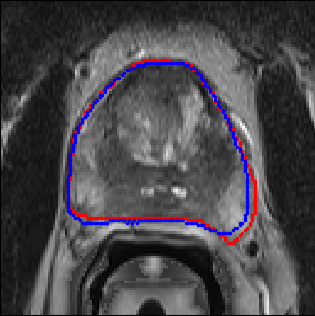} \\
    \multirow{2}{*}{ 3D Liver CT} &
    \includegraphics[width=25mm]{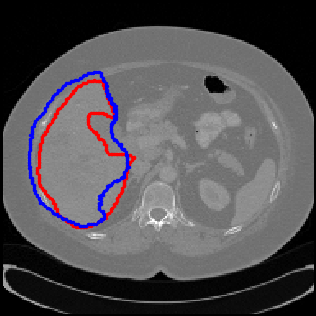} & 
    \includegraphics[width=25mm]{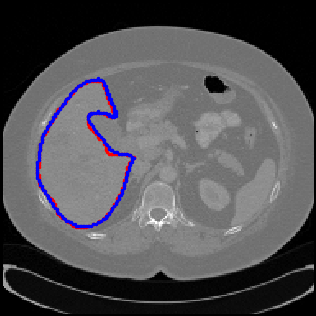} &
    \includegraphics[width=25mm]{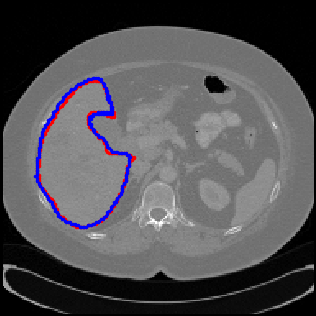} &
    \includegraphics[width=25mm]{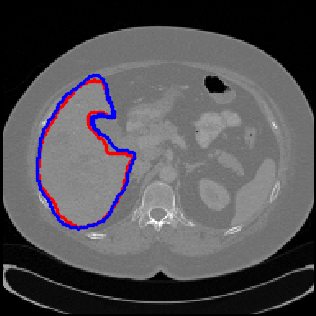} \\     
    & \includegraphics[width=25mm]{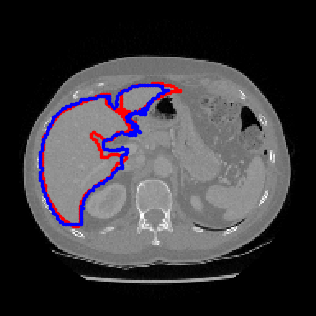} & 
    \includegraphics[width=25mm]{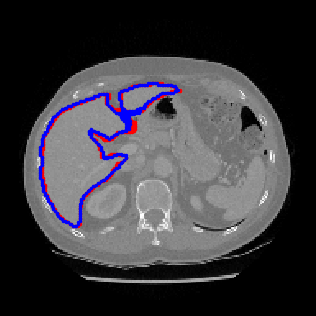} &
    \includegraphics[width=25mm]{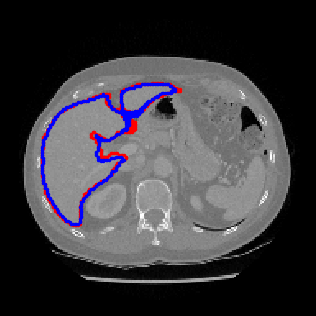} &
    \includegraphics[width=25mm]{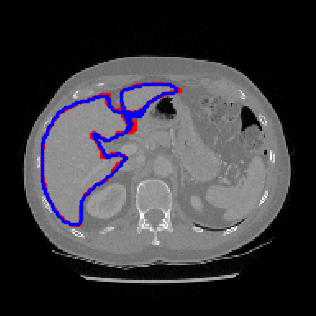} \\
    \multirow{2}{*}{ 3D Pancreas CT} &
    \includegraphics[width=25mm]{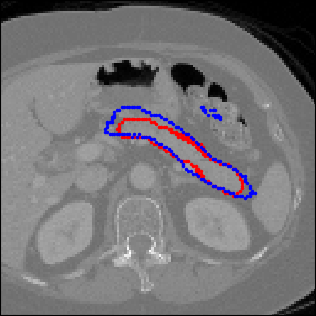} & 
    \includegraphics[width=25mm]{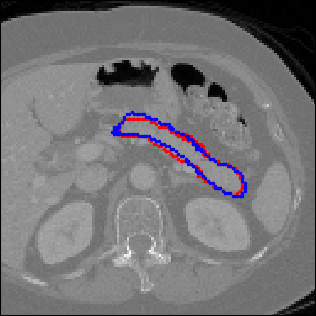} &
    \includegraphics[width=25mm]{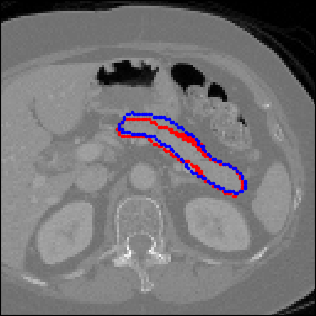} &
    \includegraphics[width=25mm]{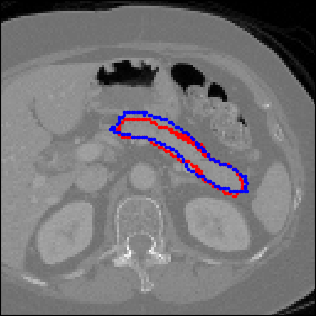} \\
    & \includegraphics[width=25mm]{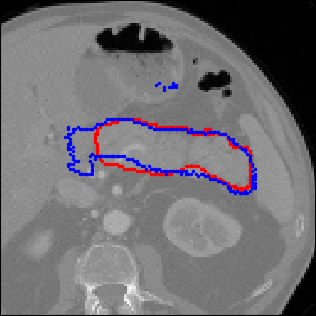} & 
    \includegraphics[width=25mm]{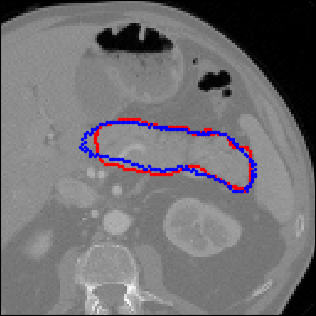} &
    \includegraphics[width=25mm]{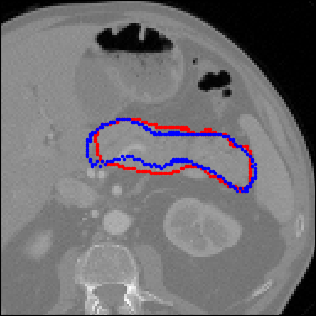} &
    \includegraphics[width=25mm]{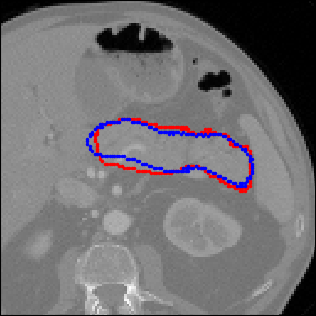} \\
   &  $f_{\text{DSC}}$ & $f^{\text{DT}}_{\text{HD}}$ & $f^{\text{ER}}_{\text{HD}}$ & $f^{\text{CV}}_{\text{HD}}$ \\
  \end{tabular}
\caption{Selected images from each dataset and the boundaries of the segmentations produced by different loss functions (in blue) and the ground-truth segmentation (in red). For the 3D images, we have shown the slice on which $f_{\text{DSC}}$ had the largest segmentation error.}
\label{fig:results}
\end{figure*}

Figure \ref{fig:results} shows example test images on which the proposed HD-based loss functions resulted in lower HD than the DSC loss. We have shown one example image from each dataset. For the 3D images, we have shown the axial slice on which the largest error occurred with the DSC loss function.

\begin{figure}[ht]
  \centering
\includegraphics[width=60mm]{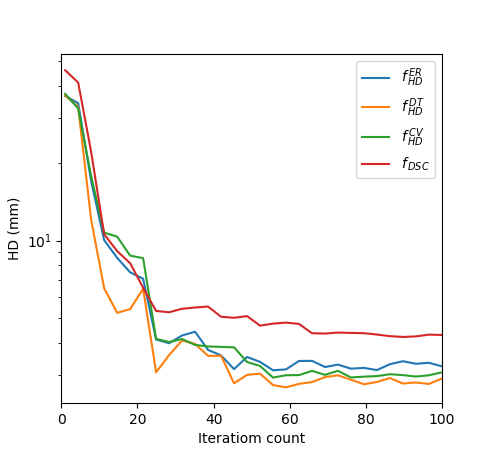}
\caption{A plot of the mean HD on the test data for the 2D TRUS images of prostate as a function of the training epoch number for different loss functions.}
\label{fig:convergence_plots}
\end{figure}

Based on the results in Table \ref{tab:overall_summary}, among the three HD-based loss functions, $f^{\text{DT}}_{\text{HD}}$ and $f^{\text{CV}}_{\text{HD}}$ resulted in lower HD than $f^{\text{ER}}_{\text{HD}}$. The difference was statistically significant on three out of the four datasets. On the other hand, the training times for $f^{\text{DT}}_{\text{HD}}$ and $f^{\text{CV}}_{\text{HD}}$ were longer than that for $f^{\text{ER}}_{\text{HD}}$. On average, $f^{\text{DT}}_{\text{HD}}$ led to the best results in terms of HD, but with training times on 3D images that were approximately twice those of  $f^{\text{ER}}_{\text{HD}}$ and $f_{\text{DSC}}$. One may speculate that training with $f^{\text{ER}}_{\text{HD}}$ may lead to segmentation performance on par with $f^{\text{DT}}_{\text{HD}}$ and $f^{\text{CV}}_{\text{HD}}$ if it is given equal training time in hours, rather than equal number of training epochs. However, our experiments showed this was not the case. As shown in Figure \ref{fig:convergence_plots}, for all loss functions the segmentation performance on the test data plateaued well before 100 epochs of training. 

There are other loss functions that could have been included in our experiments. For example, cross-entropy is also commonly used as a loss function for training deep learning-based image segmentation models. In our experiments, cross-entropy always performed worse than DSC. Using weighted cross-entropy did not significantly improve the results. For example, for the 2D prostate ultrasound data the DSC and HD achieved using weighted cross-entropy as the loss function were $0.919 \pm 0.050$ and $4.3 \pm 3.2$, respectively. For 3D CT pancreas data, the DSC and HD achieved using weighted cross-entropy as the loss function were $0.746 \pm 0.125$ and $34.0 \pm 17.3$, respectively.

We should note that although the training times for different loss functions are quite different, the test times are identical. This is because segmentation of a test image only requires a forward pass through the network and does not involve the loss function in any way. The loss functions are only used during training. On an NVIDIA GeForce GTX TITAN X, the test time for a single image for the 2D prostate ultrasound, 3D prostate MRI, 3D liver CT, and 3D pancreas CT were, respectively 0.09, 0.36, 0.45, and 0.45 seconds. These times were identical for the networks trained with any of the HD-based loss functions as well as the DSC loss.

As we mentioned in Section \ref{DistanceTransform}, with the distance transform-based approach, using a modified loss function based on the one-sided HD ($\text{Loss}_{\text{DT-OS}}$, shown in Equation \eqref{eq:dt_cost_os}) will reduce the computational cost. We performed some experiments with this loss function. Although with this loss function the training time is almost equal to that of $f_{\text{DSC}}$, the segmentation results were not very encouraging. For example on the Pancreas CT dataset, HD was $24.8 \pm 11.9$, which was statistically significantly larger than that of $\text{Loss}_{\text{DT}}$. Nonetheless, in our experiments, this low-cost loss function always reduced the HD compared with $f_{\text{DSC}}$.

Although the results with $f^{\text{ER}}_{\text{HD}}$ were slightly worse than with $f^{\text{DT}}_{\text{HD}}$ and $f^{\text{CV}}_{\text{HD}}$, it still significantly reduced HD compared with $f_{\text{DSC}}$. Moreover, it adds little computational overhead to a CNN. The likely reason for the lower performance of $f^{\text{ER}}_{\text{HD}}$ is that it is not as accurate as the other two methods in estimating HD. On some images, it can greatly underestimate HD. Our approach to estimating HD using morphological operations is a simple one. In addition to the cross-shaped structuring elements described above, we also experimented with square-shaped and cube-shaped elements (for 2D and 3D, respectively). However, the results in terms of the spread of the points in \mbox{Figure \ref{fig:morphological_operations_proof}} and also in terms of segmentation accuracy were not better than those obtained with the cross-shaped elements. Nonetheless, it may be possible to design more accurate, but still fast and simple, methods based on morphological operations proposed in \cite{mazille1989}.

For $\text{Loss}_{\text{CV}}$, the choice of the set of radius values, $R$, can be considered as a hyper-parameter, which should be chosen for each application based on the expected range of segmentation errors. As we have mentioned above, the choice of $R$ will affect both the segmentation accuracy and computational time. To give the reader a sense of this trade-off, in Table \mbox{\ref{tab:loss_cv_radius}} we have shown the results of some experiments on 2D prostate ultrasound and 3D Pancreas CT with different $R$. Some conclusions can be drawn from these results. First, as we speculated in Section \mbox{\ref{CircularConvolutions}}, no gain is achieved by using kernels of every size. This can be seen by comparing the results obtained with $R= \{ 3, 6, 9, \dots , 18 \}$ versus $R= \{ 3, 4, 5, \dots , 18 \}$ for 2D prostate ultrasound dataset and by comparing the results obtained with $R= \{ 3, 6, \dots, 15 \}$ versus $R= \{ 3, 5, \dots , 15 \}$ for 3D Pancreas CT dataset. Using more tightly-packed values of $R$ will only increase the computational cost without substantially improving the segmentation results. The results in Table \mbox{\ref{tab:loss_cv_radius}} also show that the segmentation performance is not very sensitive to the choice of $R$. In particular, for both datasets used in these experiments, all choices of $R$ resulted in much smaller HD than with $f_{\text{DSC}}(q, p)$ (as shown in Table \mbox{\ref{tab:overall_summary})}. Table \mbox{\ref{tab:overall_summary}} above also showed that the same choice of $R = \{ 3, 6, 9 \}$ led to very good results on three different datasets of 3D prostate MRI, 3D liver CT and 3D pancreas CT. This is further evidence that this method is not highly sensitive to the choice of $R$.

\begin{table*}
  \centering
\caption{The results of experiments with various choices of the parameter $R$ in $\text{Loss}_{\text{CV}}$ on two different datasets. For each dataset, the first row shows the result with our default choice of $R$ which was presented in Table \mbox{\ref{tab:overall_summary}}. The last row for each dataset shows the result obtained with $f_{\text{DSC}}$ (also from Table \mbox{\ref{tab:overall_summary}}) for easy comparison.}
\label{tab:loss_cv_radius}
\begin{tabular}{L{3.0cm} L{2.8cm} C{2.8cm} C{2.8cm} C{2.5cm}}
\hline
Dataset & $R$ & DSC & HD (mm) & Training time (h) \\ \hline 
\multirow{5}{*}{2D prostate ultrasound } 
& $R= \{ 3, 6, 9, \dots , 18 \}$  & $0.941 \pm  0.036$ & $2.7 \pm  1.8$ & 4.1 \\
& $R= \{ 3, 6, 9, \dots , 30 \}$  & $0.943 \pm  0.030$ & $2.7 \pm  1.6$ & 9.8 \\
& $R= \{ 3, 6, 9 \}$           & $0.936 \pm  0.045$ & $3.0 \pm  2.2$ & 3.8 \\
& $R= \{ 3, 4, 5, \dots , 18 \}$  & $0.940 \pm  0.035$ & $2.7 \pm  1.8$ & 12.0 \\
& $R= \{ 3, 4, 5, \dots , 30 \}$  & $0.944 \pm  0.032$ & $2.6 \pm  1.6$ & 20.4 \\
& $f_{\text{DSC}}(q, p)$         & $0.932 \pm  0.039$  & $4.3 \pm  2.8$ & 3.2 \\
\hline
\multirow{5}{*}{3D Pancreas CT } 
& $R= \{ 3, 6, 9 \}$           & $0.780 \pm  0.055$ & $21.7 \pm  11.0$ &  34 \\
& $R= \{ 3, 6, \dots, 15 \}$   & $0.790 \pm  0.056$ & $21.2 \pm  10.7$ &  51 \\
& $R= \{ 3, 6 \}$              & $0.772 \pm  0.082$ & $24.0 \pm  12.8$ &  28 \\
& $R= \{ 3, 5, \dots , 9 \}$   & $0.779 \pm  0.050$ & $21.4 \pm  10.6$ &  40 \\
& $R= \{ 3, 5, \dots , 15 \}$  & $0.788 \pm  0.061$ & $21.3 \pm  10.1$ &  63 \\
& $f_{\text{DSC}}(q, p)$       & $0.752 \pm  0.120$ & $32.1 \pm  17.0$ &  22 \\
\hline
\end{tabular}
\end{table*}

Our proposed DT-based loss function, $f^{\text{DT}}_{\text{HD}}$, is based on a weighting of the segmentation errors, where larger distance errors are weighted more strongly. This approach seems to be the opposite of what some previous studies have proposed. For example, for cell segmentation in \cite{ronneberger2015} and for prostate segmentation in \cite{anas2017}, it has been suggested that larger weights be assigned to the pixels that are closer to the boundary of the ground-truth segmentation. To test this alternate approach, we used the following loss function, which is similar to the loss function suggested in \cite{ronneberger2015}:

\begin{equation} \label{eq:dt_cost_os_previous}
\text{Loss}_{\text{DT}}^{\dagger}( q, p)= \frac{1}{\mid \Omega \mid} \sum_{\Omega} \bigg( (p-q)^2 \circ \, \exp \Big( \frac{-d_p^2} {2 \sigma^2} \Big) \bigg)
\end{equation}

Our observations show that although this loss function may slightly improve the DSC on some datasets, in general it has no significant positive effect on HD, which is the focus of this work. For example, with the above loss function (with an added DSC loss term as in Equation \eqref{eq:hd_loss_general}), on the 2D TRUS prostate data we achieved DSC and HD of $0.938 \pm  0.035$  and $4.0 \pm  2.7 \text{ mm}$, respectively. Paired t-tests showed that HD was significantly larger than our three HD-based loss functions and DSC was not significantly different from those obtained with other loss functions on this dataset. Therefore, compared with our proposed loss functions, assigning larger weights to the pixels closer to the ground-truth boundary harms the segmentation performance in terms of HD. It is worth pointing out that the main challenge in \cite{ronneberger2015} was to segment the boundaries of the object of interest (cells), which could justify a loss function as in Equation \eqref{eq:dt_cost_os_previous}. Moreover, unlike \cite{ronneberger2015, anas2017}, our loss functions include a DSC loss term, which means that the comparison of our results with those studies is not quite fair.

\begin{figure}[!ht]
  \centering
\begin{tabular}{c}
    \includegraphics[width=60mm]{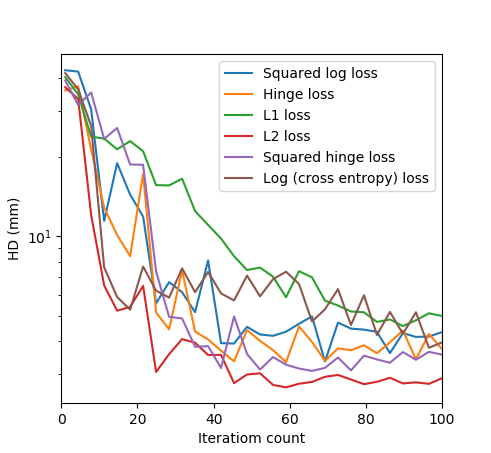} \\
    \includegraphics[width=60mm]{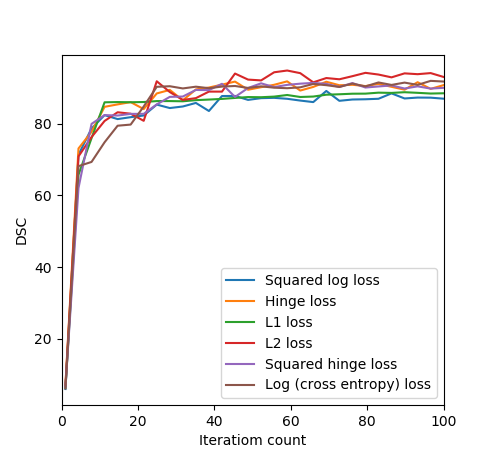} \\
  \end{tabular}  
\caption{Plots of Dice Similarity Coefficient and Hausdorff Distance on the 2D TRUS images of the prostate for different formulations of the distance transform-based loss function.}
\label{fig:different_inner_losses}
\end{figure}

Another distinct aspect of our proposed HD-based loss functions is that they are based on the squared difference of the probability maps, i.e., $(p-q)^2$, whereas medical image segmentation methods typically work with the cross-entropy. Our choice of the $\ell_2$-norm was motivated by the results reported in some recent studies \cite{janocha2017, lee2015, tang2013}. These studies have shown that loss functions based on hinge loss, squared hinge loss, and $\ell_1$ and $\ell_2$ norms may lead to superior results in different deep learning models. Inspired by these studies, and because none of them had considered the application of image segmentation, we conducted a set of experiments to examine the usefulness of these formulations in our application. Figure \ref{fig:different_inner_losses} shows an example of our observations. In this experiment, we replaced the squared difference term $(p-q)^2$ in the DT-based loss function (Equation \eqref{eq:dt_cost}) with some of the alternatives proposed in \cite{janocha2017, lee2015, tang2013}. As can be seen in this figure, $\ell_2$ loss, hinge loss, and squared hinge loss all perform better than the cross-entropy loss, which is widely used in CNN-based image segmentation methods. It was based on such observations that we built our HD-based loss functions (Equations \eqref{eq:dt_cost}, \eqref{eq:er_cost}, and \eqref{eq:cv_cost}) upon the squared difference, $(p-q)^2$. Overall, our observation are in line with those reported in \cite{janocha2017}. However, we observed that the $\ell_2$ loss gives slightly better results than the hinge and squared hinge losses and that the $\ell_1$ loss is not very poor either, whereas the authors of \cite{janocha2017} found that the $\ell_1$ loss was very poor and the hinge losses were slightly better than the $\ell_2$ loss. We think that the most likely reason for these differences is the extra DSC loss term in our work. Moreover, our image segmentation problem is quite different from the applications considered in \cite{janocha2017}.

Overall, the results reported in Table \ref{tab:overall_summary} are close to or better than the results reported by many recent studies. On the 2D TRUS prostate image data, our results in terms of HD  are much better than those reported in \cite{nouranian2015, nouranian2016} on the same dataset, where the authors have reported HD of above $5 \text{ mm}$ using two different methods. For prostate segmentation in 3D MRI, most studies have only reported DSC. Some studies have reported the 95th-percentile of HD within an image \cite{milletari2016,salimi2018,brosch2018}. Note that this is different from the \textit{inter-patient} 90th-percentile of HD that we have reported in this paper. The mean of the 95th-percentile intra-patient HD reported in a recent comparison of several state of the art methods is in the range $4.9 - 7.6 \text{ mm}$ \cite{brosch2018}, whereas this quantity computed on the test data with the model trained using $f^{\text{DT}}_{\text{HD}}(q, p)$ in our work is $4.70 \pm  0.97 \text{ mm}$. Incidentally, the lowest HD in the comparison published in \cite{brosch2018} was achieved by a method from our own group that has not been published yet. For liver segmentation in CT, the reported values of HD  vary greatly between $24 \text{ mm}$ and $119 \text{ mm}$ \cite{christ2016}. Our best result, as can be seen in Table \ref{tab:overall_summary} is $25.1 \text{ mm}$. For pancreas segmentation, most studies only report DSC and the mean or root-mean-square of surface distance. A recent study reported values of HD in the range $17.7 - 22.2 \text{ mm}$ \cite{roth2018b}, compared with our best result of $21.3 \text{ mm}$. Pancreas segmentation is considered to be very challenging and most studies have reported a DSC of below $0.80$. Recently some works have achieved DSC values of well above $0.80$ \cite{roth2018,zhou2017}. We should note, however, that those studies have used more elaborate machinery such as a separate model to identify the location of the pancreas or iterative refinement strategies. For example, the work that has reported mean HDs as low as $17.7 \text{ mm}$ is a multi-stage method that uses multiple CNN models and other machine learning methods to localize the pancreas, identify boundary cues, and aggregate segmentation cues \cite{roth2018b}. We should stress that the goal of the present study was not to achieve the state of the art results in segmentation of different organs and imaging modalities; one could always achieve better results by fine-tuning the network structure or employing a more sophisticated methodology that is tailored to the specific organ and imaging modality. Our experiments were intended to show that the proposed loss functions could lead to a significant reduction in HD. That is why we have adopted standard segmentation models and training procedures in order to eliminate, as much as possible, other confounding factors and study the effectiveness of the proposed loss functions.

We further compared our proposed loss functions on the test data of the PROMISE12 challenge \mbox{\cite{litjens2014b}}. In this experiment, we trained our CNN model with different loss functions on the 50 training images provided by this challenge and tested on the 30 test images for which the true segmentation is only known to the challenge organizers. A summary of the results of this experiment is presented in Table \mbox{\ref{tab:promise12}}. The trends observed in this Table are similar to those in Table \mbox{\ref{tab:overall_summary}}. The DSC scores achieved on this dataset are better than those in Table \mbox{\ref{tab:overall_summary}} and they are very similar for all four loss functions. Compared with $f_{\text{DSC}}$, all three proposed HD-based loss functions have substantially reduced the mean, standard deviation, and the maximum of HD95. The reduction in HD for 3D prostate MRI data in Table \mbox{\ref{tab:overall_summary}} was in the range $18-23 \%$, whereas the reduction in HD95 in \mbox{\ref{tab:promise12}} was in the range $10-15 \%$. This is, at least in part, because HD95 ignores the top $5 \%$ with the largest surface distance error. Therefore HD95 does not reflect the very largest segmentation errors that HD represent. We cannot compute the HD for this dataset because we do not have the ground-truth. Nonetheless, these results show that our proposed method is capable of reducing not only the very largest segmentation error but also to consistently reduce large segmentation errors as quantified by HD95. Compared with other recently-published papers on the same dataset, our achieved HD95 values are significantly better. Two recently-published methods evaluated on the same dataset have been included in Table \mbox{\ref{tab:promise12}}. As we pointed out above, a recent study compared 10 state of the art methods on this dataset and reported the HD95 values in the range $4.9 - 7.6 \text{ mm}$ \mbox{\cite{brosch2018}}, whereas our HD-based loss functions resulted in HD values as low as 4.26.

\begin{table*}
\caption{A summary of the comparison of the proposed loss functions on the test data from the PROMISE12 challenge \mbox{\cite{litjens2014b}}. N.R. stands for ``not reported".}
\label{tab:promise12}
\begin{tabular}{L{4.5cm} C{3.5cm} C{3.5cm} C{3.5cm} }
\hline
 Loss Function & DSC & HD95 (mm) & Max. of HD95 (mm) \\ \hline 
$f_{\text{DSC}}(q, p)$   & \boldmath$0.908 \pm  0.032$ & $5.00 \pm  2.16$ & 11.3 \\
$f^{\text{DT}}_{\text{HD}}(q, p)$   & $0.902 \pm  0.026$ & $4.28 \pm  1.05$ & 7.8 \\
$f^{\text{ER}}_{\text{HD}}(q, p)$   & $0.904 \pm  0.023$ & $4.48 \pm  1.46$ & 8.8 \\
$f^{\text{CV}}_{\text{HD}}(q, p)$   & $0.902 \pm  0.025$ & \boldmath$4.26 \pm  1.03$ & \boldmath$7.6$ \\
Yu et al., 2017, \cite{yu2017}   & $0.894$ & $5.54$ & N.R. \\
Brosch et al., 2018, \cite{brosch2018}  & $0.905$ & $4.94$ & N.R. \\
\hline
\end{tabular}
\end{table*}

We also applied our methods on the MRI brain segmentation data from the iSeg-2017 challenge \mbox{\cite{wang2019}}. The goal of this challenge is to segment infant brain MR images into four classes: 1) white matter (WM), 2) gray matter (GM), 3) cerebrospinal fluid (CSF), and 4) background. The challenge organizers allow a maximum of two submissions per team. Because based on Table \mbox{\ref{tab:overall_summary}} $f^{\text{DT}}_{\text{HD}}$ and $f^{\text{ER}}_{\text{HD}}$ were, overall, the best and worst of our three HD-based loss functions, we evaluated these two loss functions on this challenge. Our results have been shown in Table \mbox{\ref{tab:iSeg2017}}. As a comparison with another method, we have included the results obtained by the recently published method in \mbox{\cite{hashemi2019}}, which at the time of our participation in this challenge (March 2019) has achieved the highest Dice score on all three structures among all participating teams. To also compare our other two loss functions ($f_{\text{DSC}}$ and $f^{\text{CV}}_{\text{HD}}$) we performed a five-fold cross-validation on the training images of this challenge. The results of this experiment have also been included in Table \mbox{\ref{tab:iSeg2017}}.

As can be seen from this table, $f^{\text{DT}}_{\text{HD}}$ and $f^{\text{ER}}_{\text{HD}}$ achieve a lower HD than the method of \mbox{\cite{hashemi2019}} on all three structures, with one exception of loss function $f^{\text{ER}}_{\text{HD}}(q, p)$ on CSF segmentation). Our best performing loss function ($f^{\text{DT}}_{\text{HD}}(q, p)$) has reduced HD compared with \mbox{\cite{hashemi2019}} by approximately $1.5 \%$, $35 \%$, and $4.2 \%$, respectively on CSF, GM, and WM, respectively. Overall, compared with all participating teams in this challenge, our best achieved HD was ranked 1st in CSF segmentation, 3rd in GM segmentation, and 6th in WM segmentation. We should note that we used the basic 3D U-Net that we had used for segmentation of the other datasets in this study, without any modifications. On the other hand, other participating teams have developed methods specifically for this challenge. For example, the leading method in \mbox{\cite{hashemi2019}} has used a CNN-based patch-level training and prediction aggregation method specially designed for this brain MRI segmentation task. A better way of assessing the effect of our proposed HD-based loss functions is to compare with the results obtained with $f_{\text{DSC}}$ in our five-fold cross-validation experiments reported in the same Table. Compared with $f_{\text{DSC}}$, our best performing loss function, $f^{\text{DT}}_{\text{HD}}$, reduces HD by $15 \%$, $35 \%$, and $23 \%$, respectively on CSF, GM, and WM, respectively, which represent substantial improvements. The other two HD-based loss functions, $f^{\text{ER}}_{\text{HD}}$ and $f^{\text{CV}}_{\text{HD}}$, also reduce the HD (compared with $f_{\text{DSC}}$) by $8 \%$ to $33 \%$ while achieving DSC values that are very close to or better than $f_{\text{DSC}}$.

\begin{table*}
\caption{A summary of the comparison of the proposed loss functions on the test data from the iSeg-2017 challenge.}
\label{tab:iSeg2017}
\begin{tabular}{L{5.0cm} L{3.0cm} | C{1.0cm} C{1.0cm} | C{1.0cm} C{1.0cm} | C{1.0cm} C{1.0cm} }
\hline
& & \multicolumn{2}{c}{CSF} & \multicolumn{2}{c}{GM} & \multicolumn{2}{c}{WM} \\ \hline
 Experiment & Loss Function & DSC & HD & DSC & HD & DSC & HD  \\ \hline 
\multirow{3}{*}{\parbox{4.5cm}{Evaluated on the test data of iSeg2017}} & $f^{\text{DT}}_{\text{HD}}(q, p)$ & 0.945 & \boldmath$8.720$  & 0.911 & \boldmath$6.212$ & 0.890 & \boldmath$6.812$ \\ 
& $f^{\text{ER}}_{\text{HD}}(q, p)$ & 0.947 & 9.434  & 0.915 & 6.832 & 0.894 & 6.988 \\
& Hashemi et al., 2019 \cite{hashemi2019} & \boldmath$0.960$ & 8.850  & \boldmath$0.926$ & 9.557 & \boldmath$0.907$ & 7.104 \\
\hline
\multirow{4}{*}{\parbox{4.5cm}{Evaluated on the training data of iSeg2017 via five-fold cross-validation}} & $f_{\text{DSC}}(q, p)$ & 0.950 & 10.224  & 0.914 & 9.320 & 0.905 & 8.802 \\
& $f^{\text{DT}}_{\text{HD}}(q, p)$ & 0.942 & \boldmath$8.725$  & 0.910 & \boldmath$6.101$ & 0.894 & \boldmath$6.816$ \\ 
& $f^{\text{ER}}_{\text{HD}}(q, p)$ & 0.949 & 9.408  & 0.910 & 6.845 & 0.899 & 6.983 \\
& $f^{\text{CV}}_{\text{HD}}(q, p)$ & \boldmath$0.951$ & 8.733  & \boldmath$0.922$ & 6.225 & \boldmath$0.908$ & 6.854 \\
\hline
\end{tabular}
\end{table*}

As we have argued throughout the paper, minimizing HD directly can be tricky and counter-productive. This is why we have proposed relaxed loss functions that are based on HD, instead of minimizing HD directly. Moreover, we augmented our HD-based loss functions with DSC loss, which is based on the amount of overlap between the ground-truth and estimated segmentation maps. Nonetheless, one would be curious to know how the proposed HD-based loss functions would perform without the added DSC loss term. We performed experiments to empirically understand the training of CNN segmentation methods with these loss functions. We observed that it was possible to train CNN segmentation methods with these loss functions. However, this required a more careful tuning of the learning rate and using a much smaller learning rate in the first few training epochs. Moreover, we observed that the segmentation results were always better when the DSC loss term was included.

Here we briefly summarize the results obtained on two of our datasets. For the 2D prostate ultrasound data, the HD values achieved using $f^{\text{DT}}_{\text{HD}}$, $f^{\text{ER}}_{\text{HD}}$, and $f^{\text{CV}}_{\text{HD}}$ without the DSC loss term were, respectively, $2.9 \pm  1.8$, $3.3 \pm  2.4$, and $2.9 \pm  2.2$ and the DSC values were, respectively, $0.930 \pm  0.038$, $0.921 \pm  0.048$, and $0.923 \pm  0.040$. These results are slightly worse than those reported in Table \mbox{\ref{tab:overall_summary}} for these methods with the added DSC loss term. Nonetheless, it is interesting to note that these HD values are still much better than $4.3 \pm  2.8$ achieved with $f_{\text{DSC}}$ as shown in Table \mbox{\ref{tab:overall_summary}}. For the 3D pancreas CT data, the HD values achieved using $f^{\text{DT}}_{\text{HD}}$, $f^{\text{ER}}_{\text{HD}}$, and $f^{\text{CV}}_{\text{HD}}$ without the DSC loss term were, respectively, $22.9 \pm  13.8$, $27.6 \pm  13.4$, and $22.7 \pm  12.0$ and the DSC values were, respectively, $0.779 \pm  0.058$, $0.750 \pm  0.068$, and $0.772 \pm  0.054$. These results also show that better results can be obtained by including the DSC loss term. Nonetheless, these HD values are still much smaller than $32.1 \pm  17.0$ obtained with $f_{\text{DSC}}$ as shown in Table \mbox{\ref{tab:overall_summary}}.

\begin{figure}[ht]
  \centering
\includegraphics[width=80mm]{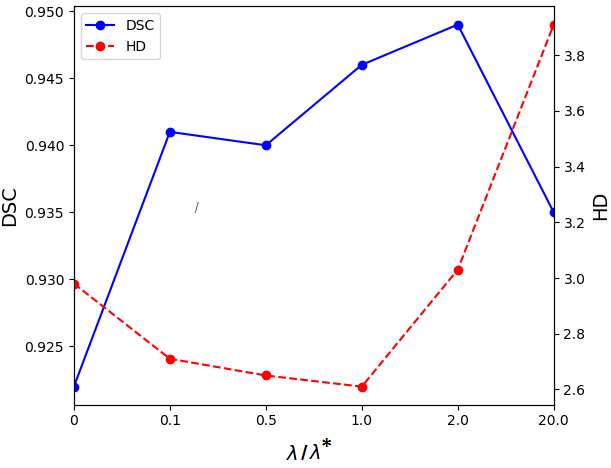}
\caption{The change in the segmentation performance on the 2D prostate ultrasound data using loss function $f^{\text{DT}}_{\text{HD}}$ with different values of $\lambda$. The horizontal axis is a function of $\lambda / \lambda^*$, where $\lambda^*$ is our default setting which is chosen such that the HD-based and DSC loss terms have equal weights.}
\label{fig:lambda}
\end{figure}

\begin{figure}[ht]
  \centering
\includegraphics[width=80mm]{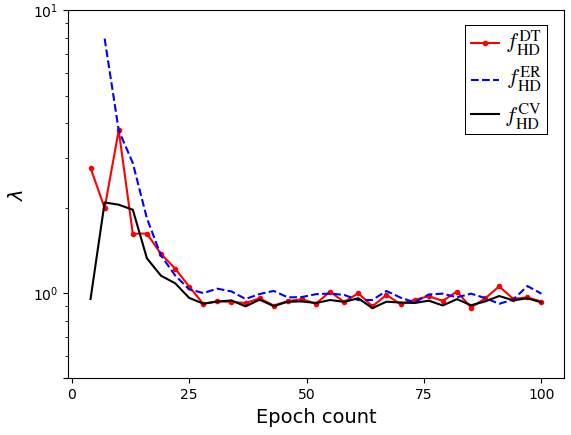}
\caption{Change in the value of $\lambda$ with training epochs for the 3D liver CT dataset. The values of $\lambda$ for the three loss functions have been normalized such that the value at the end of training is approximately equal to one.}
\label{fig:lambda_trend}
\end{figure}

Using the HD-based loss functions alone is equivalent to setting $\lambda=0$ in Eq. \mbox{\ref{eq:hd_loss_general}}. Using the DSC loss alone, i.e., $f_{\text{DSC}}$ corresponds to a very large $\lambda$. As we mentioned above, in our experiments we update the value of $\lambda$ such that the two loss terms (i.e., the HD-based loss term and the DSC loss term) have equal weight. In Figure \mbox{\ref{fig:lambda}} we have shown the effect of changing this weighting on the example of training with $f^{\text{DT}}_{\text{HD}}$ on the 2D prostate ultrasound data. This figure shows that for this specific case this choice is close to optimal and that choosing $\lambda$ such that the ratio of the DSC loss term to the HD-based loss term is in the range $[0.1, 2]$ leads to good segmentation results in terms of both HD and DSC. Figure \mbox{\ref{fig:lambda_trend}} shows the change in the value of $\lambda$ over training epochs for the three HD-based loss functions on the 3D liver CT data. We observed similar trends on the other datasets. The values of $\lambda$ have been normalized such that the value at the end of training is approximately equal to one, so that the curves can be shown on the same plot. The overall trend is that the value of $\lambda$ in the early training epochs is larger. This is because at the start of training there are many false positives far from the segmentation boundaries, which increases the HD-based loss term.

We further tried training with the ``exact HD" as the loss function, instead of our proposed relaxed HD-based loss functions. In these experiments, we tried training models using Equation \mbox{\eqref{eq:dt3}} as the loss function for 2D and 3D datasets.  However, the models never converged to a meaningful state regardless of the learning rate. This is very much expected because HD aims at minimizing the error at one single point with the largest error. We have presented arguments against using HD as a loss function in Section \mbox{\ref{Introduction}} and reviewed some of the computer vision literature in this regard. Moreover, with randomly-initialized deep learning models, minimizing only the largest error cannot be justified and our observations confirm this.

We have successfully tested our method on five different datasets which include organs of different sizes and shapes in different imaging modalities. Nonetheless, we cannot claim that our method will be successful in all medical image segmentation tasks. Indeed, the range of anatomical shapes and the level of detail in medical image segmentation is very wide. Hence, application of our loss functions to other organs may need modifications. For example, in some applications such as vessel segmentation, recent studies have used additional modules such as probabilistic graphical models on top of CNNs to achieve good results \mbox{\cite{fu2016,shin2018}}. In such applications, our proposed methods might need application-specific modifications. A comprehensive experimental study or review of the range of applications for our proposed method and a discussion of all application-specific issues is beyond the scope of this paper.

\section{Conclusion}
\label{Conclusions}

Our results show that all three proposed HD-based loss functions can lead to statistically significant reductions in HD in the segmentation of 2D and 3D medical images of different imaging modalities. Therefore, the proposed methods can be very useful in applications such as multimodal image registration where large segmentation errors can be very harmful. To the best of our knowledge, none of the three methods proposed in this work for estimating HD and the three loss functions proposed for training segmentation algorithms have appeared in previous publications. The distance transform-based loss, $f^{\text{DT}}_{\text{HD}}$, is the most intuitive of the three formulations and leads to very good results, but it also substantially increases the computational load. The loss based on morphological erosion, $f^{\text{ER}}_{\text{HD}}$, is computationally less expensive, but not as effective as the other two losses. The loss based on convolutions with kernels of increasing sizes, $f^{\text{CV}}_{\text{HD}}$, leads to very good results and it can be computationally much less demanding than the distance transform-based loss if one properly limits the maximum size and the number of convolutional kernels that are used. Overall, the decision on which of the three loss functions to use depends on the application. For example, for 2D images the cost of computing the distance transforms is not substantial and one may use $f^{\text{DT}}_{\text{HD}}$. For 3D images, $f^{\text{ER}}_{\text{HD}}$ and $f^{\text{CV}}_{\text{HD}}$ may be better options, with $f^{\text{CV}}_{\text{HD}}$ offering better segmentation performance at the cost of longer training times.

To the best of our knowledge, this is the first work to aim at reducing HD in medical image segmentation. The methods presented in this paper may be improved in several ways. Faster implementation of the HD-based loss functions and more accurate implementation of the loss function based on morphological erosion would be useful. Moreover, extension of the methods for other applications such as vessel segmentation could also be pursued.

\section*{Acknowledgment}

This project is funded by the Natural Science and Engineering Research Council of Canada (NSERC), the Canadian Institutes of Health Research (CIHR), and the Prostate Cancer Canada (PCC). We would like to thank the support from the Charles Laszlo Chair in Biomedical Engineering held by Professor S. Salcudean.

\ifCLASSOPTIONcaptionsoff
  \newpage
\fi

\bibliographystyle{IEEEtran}
\bibliography{bib3}

\end{document}